\documentstyle[prb,aps,epsfig,prbbib]{revtex}
\begin{document}
\draft
\twocolumn[\hsize\textwidth\columnwidth\hsize\csname  
@twocolumnfalse\endcsname
\title{Evolution from BCS superconductivity to Bose-Einstein condensation:
Current correlation function in the broken-symmetry phase}
\author{N. Andrenacci$^{1,2}$, P. Pieri$^1$, and G.C. Strinati$^1$}
\address{(1)  Dipartimento di Fisica, UdR INFM
\\ Universit\`{a} di Camerino, I-62032 Camerino, Italy
\\ (2) Institut de Physique, Universit\'e de Neuchatel 
\\CH-2000 Neuchatel,  Switzerland}
\date{\today}
\maketitle
\hspace*{-0.25ex}

\begin{abstract}

We consider the current correlation function for a three-dimensional system
of fermions embedded in
a homogeneous background and mutually interacting via an attractive
short-range potential, \emph{below\/}
the (superconducting) critical temperature.
Diagrammatic contributions in the broken-symmetry phase are identified,
that yield for the (wave-vector and
frequency dependent) current correlation function the fermionic
BCS approximation in the
weak-coupling limit and the bosonic Bogoliubov approximation in the
strong-coupling limit (whereby composite bosons form as bound-fermion pairs).
The temperature dependence of the superfluid density (from the BCS
exponential behavior at weak coupling
to a power-law behavior at strong coupling) and the form of the
Pippard-like kernel at zero temperature
are explicitly obtained from weak to strong coupling.
Quite generally, it is shown that the Pippard-like kernel is the sum of a
local (London-like) term and of a
nonlocal component, the local term being dominant in the 
strong-coupling limit and the nonlocal component in the BCS
(weak-coupling) limit.
It is also shown that the range of the nonlocal component is determined by
the coherence length measuring
the spatial correlations of the amplitude of the order parameter, namely,
the correlations \emph{among\/}
different Cooper pairs (or composite bosons), rather than between the fermionic
partners of a given pair.
In addition, a prescription is provided for mapping the fermionic onto the
bosonic diagrammatic theories
in the broken-symmetry phase, thus complementing what already done in
the normal phase.
\end{abstract}
]
\narrowtext
\section{Introduction}
Already from London and Pippard phenomenological approaches to
superconductivity, the current
response to an external magnetic field played a key role in accounting for
the Meissner effect.
The Meissner effect was, in fact, later demonstrated explicitly within the
microscopic BCS approach,
by examining the behavior of the (static) current correlation function
\cite{Schrieffer,FW}.

The BCS approach as it stands applies to weak-coupling superconductors,
with largely overlapping
Cooper pairs.
Recently, however, it has become of interest to study the evolution of a
superconducting system from
weak to strong coupling, with the strong-coupling limit corresponding
to a system
of nonoverlapping (composite) bosons
\cite{Randeria-90,Haussmann,PS-94,PS-96,Levin,Zwerger,Pi-S-98}.
The physical motivation to these works stems essentially from cuprate
superconductors, whereby the small
value of the (superconducting) coherence length as well as the presence of
a pseudogap above the critical
temperature in the underdoped region \cite{Ding,Loeser} have suggested the
possible relevance
of a \emph{crossover} scenario from a weak-coupling limit for overdoped
samples (with Cooper pairs
forming and condensing at the critical temperature within a BCS
description) toward a strong-coupling
limit for underdoped samples (with preformed (composite) bosons existing
above the superconducting
critical temperature and Bose-Einstein (BE) condensing below it).
For cuprate superconductors, it seems actually likely that the pairing
corresponds to an \emph{intermediate\/}
regime between overlapping Cooper pairs and nonoverlapping composite bosons.
More generally, the possible occurrence of a crossover from BCS to BE has
been emphasized for other
classes of ``unconventional'' superconductors as well \cite{Uemura}.

The first discussion of the crossover from BCS to BE can be found in the
pioneering paper by Eagles
dealing with superconductivity in low-carrier doped superconductors
\cite{Eagles}.
A systematic approach to this problem was then provided by Leggett
\cite{Leggett}, who showed that for a
fermionic system with an attractive interaction at zero temperature a
\emph{smooth crossover\/} from a BCS
ground state of overlapping Cooper pairs to a condensate of composite
bosons occurs as the strength
of the attraction increases.
A few years later the problem was reconsidered by Nozi\`{e}res and
Schmitt-Rink \cite{NSR}, who extended
the approach to finite temperatures by using diagrammatic methods, always
finding a \emph{smooth crossover\/}
between the two limits.

Several quantities have so far been calculated within the BCS-BE
crossover approach, including the critical
temperature and chemical potential \cite{Haussmann-2}, the coherence
length(s) at zero temperature~\cite{PS-96},
the spectral function above~\cite{PPSC} and below~\cite{leo} the critical 
temperature, and the effect of an external magnetic field on 
the pseudogap temperature~\cite{dennis}. 
No systematic study yet exists, however, for the current correlation
function and derived quantities,
such as the (temperature dependence of the) superfluid density and the
Pippard kernel.
Purpose of this paper is to provide a detailed calculation of these
quantities, by relying on controlled
approximations in the two (weak- and strong-coupling) limits, which allow
us to examine specifically how
the response of the original Fermi system can be interpreted in terms of
the response of an effective Bose
system in the strong-coupling limit.

We shall rely on a diagrammatic approach that selects the relevant
contributions to the (wave-vector and
frequency dependent) current correlation function in the weak- and
strong-coupling limits separately, and
then sum these contributions to obtain the current correlation function for
all couplings.
The contributions selected in the strong-coupling limit will appear as 
fluctuation contributions when extrapolated to the weak-coupling limit.
In this paper, we will consider a Fermi system with an
attractive (point-contact) interparticle interaction embedded in a
three-dimensional continuum, with an isotropic ($s$-wave) gap function.
The ensuing diagrammatic theory in the \emph{broken-symmetry phase\/} will
be thus simplified considerably,
yet reproducing the desired features in both limits, in a similar fashion
to what was done in the normal phase
above the critical temperature \cite{{Pi-S-98},SPL}.
No lattice effects nor the occurrence of an angular dependent ($d$-wave)
gap function will accordingly be taken
into account, deferring their study to future work.
The present approach should be regarded as a necessary preliminary
step toward the development of a theory of
the current response function over the whole crossover range for more 
realistic models.

As far as the current correlation function of interest is concerned, the
two (weak- and strong-coupling) limits
will be described by the BCS approximation (for fermions) and the
Bogoliubov approximation (for bosons),
respectively.
Again, these represent the simplest approximations which can be conceived
to describe fermionic and bosonic
systems, respectively, in the broken-symmetry (superfluid) phase, and share
analogous self-consistent constructions
to generate the relevant excited states within a temperature-dependent
mean-field approximation.
Albeit the Bogoliubov approximation for bosons is known to suffer from
several shortcomings \cite{Luban,Bassani-GCS}
(which could be overcome, at least in principle, by considering more
involved approximations for condensed
bosons \cite{HM}), we shall anyhow 
restrict ourselves to this approximation for the bosonic limit, since we are 
ultimately interested in obtaining a reasonable
strong-coupling limit while following the evolution of the
\emph{fermionic\/} current correlation function from
weak to strong coupling.
In this respect, we should mention that the Bogoliubov approximation has been
recently adopted to get the response function near zero temperature for a 
Bose gas in the context of the BE condensation of ultracold atomic dilute 
gases~\cite{heiselberg}.

While the BCS approximation to the current correlation function is well
established in weak coupling \cite{Schrieffer,FW}, recovering
the Bogoliubov results for (composite) bosons from an originally fermionic
theory is altogether nontrivial and
will therefore be considered at some length in this paper.
Specifically, we shall identify the fermionic diagrams (over and above the
standard BCS ``bubble'' contribution)
which reproduce both the \emph{longitudinal\/} and \emph{transverse\/}
contributions to the (wave-vector and
frequency dependent) Bogoliubov current correlation function,
namely, the ``ladder'' diagrams known for making the BCS
result conserving \cite{Schrieffer} and
the Aslamazov-Larkin (AL) type diagram(s) \cite{AL} here extended to the
broken-symmetry phase, in the order.
Since these diagrams can be evaluated analytically in the strong-coupling
limit, a detailed account of
this derivation will be provided in what follows.
In this context, the prescription for mapping the fermionic onto the
bosonic diagrammatic theories in the broken-symmetry
phase will also be provided, in an analogous fashion to what already
achieved in the normal phase \cite{Pi-S-98}.

Besides providing a formal theory that bridges weak- and strong-coupling
approximations to the current correlation
function, the main physical results of this paper concern, in particular,
the temperature dependence of the superfluid
density (from the BCS exponential behavior at weak coupling to a power-law
behavior at strong coupling) and the
form of the Pippard-like kernel at zero temperature~\cite{Pippard}.
The contributions of the BCS and AL diagrams to the temperature dependence
of the superfluid density will be numerically calculated for various couplings
from weak to strong, and their relative importance assessed.  
[These calculations require only the \emph{static\/} contribution to the
current correlation function, so that the analytic continuation from 
Matsubara to real frequencies is not required.]
It will be also shown that the Pippard-like kernel is the sum of a local
(London-like) term and of a nonlocal component,
where only the local term survives in the strong-coupling limit,
thus correctly reproducing the result for
non-interacting (composite) bosons, while only the nonlocal component
survives in the BCS (weak-coupling) limit, as
expected.
It will be further shown that the characteristic length that controls the
range of the nonlocal component coincides
(for a clean system) with the coherence length $\xi_{{\rm phase}}$ introduced in
Ref.~\onlinecite{PS-96} as a measure of the
spatial correlations of the amplitude of the order parameter, over the
whole crossover range from weak to strong
coupling.
The length $\xi_{{\rm phase}}$ relates the correlations \emph{among\/} 
different
Cooper pairs or between composite bosons, while correlations between fermionic 
partners in a pair is associated with the different lenght scale 
$\xi_{\rm{pair}}$.~\cite{PS-96}
The quantity $\xi_{{\rm phase}}$, in turn, coincides with the Pippard 
coherence length $\xi_0$ (and with $\xi_{\rm{pair}}$) in the
weak-coupling limit, decreases as the
coupling is increased from the weak- toward the intermediate-coupling
regime, and eventually starts again to increase
upon approaching the strong-coupling (bosonic) limit where 
$\xi_{{\rm phase}} \gg \xi_{{\rm pair}}$
\cite{PS-96} (see also Ref.~\onlinecite{btcc} for an extension to the lattice 
case).

The plan of the paper is as follows.
In Section II we consider the current correlation function within the
(conserving) BCS approximation (i.e., bubble
plus ladder diagrams) and show that, in the strong-coupling limit, it maps
onto the \emph{longitudinal\/} part of the
current correlation function for (composite) bosons within the Bogoliubov
approximation.
In Section III we identify, in addition, the contribution to the fermionic
current correlation function that maps onto
the \emph{transverse\/} part of the current correlation function for
(composite) bosons within the Bogoliubov approximation.
In Section IV numerical results for the superfluid density vs temperature
as well as for the Pippard-like kernel at zero temperature are presented, over
the whole range from weak to strong coupling.
Section V gives our conclusions.
The Appendices discuss more technical material.
Specifically, in Appendix A the connection between the particle-particle
ladder in the broken-symmetry phase and the
Bogoliubov propagators is obtained;
in Appendix B the mapping between the fermionic and bosonic diagrammatic
structures in the broken-symmetry phase is
established;
in Appendix C the form of the wave-vector and frequency dependent
current correlation function for bosons within
the Bogoliubov approximation is discussed, with emphasis on the ensuing
form of the Pippard-like kernel; and in Appendix D the BCS bubble is examined 
in a novel way as to extract its kernel for all couplings and not only in the
weak-coupling limit.

\section{Current correlation function within the BCS approximation in the
strong-coupling limit}
In this Section, we reconsider the well-known BCS approximation for the
current correlation
function \cite{Schrieffer}, in the form which fulfills conservation laws
\cite{Baym}.
We will specifically show that \emph{in the strong-coupling limit\/} of the
fermionic attraction (and at temperatures small compared with the dissociation
energy of composite bosons) this approximation maps onto the 
\emph{longitudinal\/} part of the current correlation
function for bosons within the Bogoliubov approximation.
This result, which is \emph{per se\/} nontrivial, requires us to
search for
additional contributions to the current correlation function which, in the
strong-coupling
limit, reproduce the \emph{transverse\/} part of the current correlation
function for
bosons within the Bogoliubov approximation, as discussed in the next Section.

Quite generally, the current correlation function at finite temperature in
the imaginary
time (Matsubara) representation reads \cite{FW}:
\begin{equation}
\chi_{\gamma, \gamma'}({\mathbf r}\tau,{\mathbf r'}\tau') =  - 
\langle T_{\tau}[j_{\gamma}({\mathbf r}\tau)  j_{\gamma'}({\mathbf r'}\tau')]
\rangle
\label{current-current-definition}
\end{equation}
where $0 \leq \tau,\tau' \leq \beta$ ($\beta=1/(k_{B} T)$ being the inverse
temperature and
$k_{B}$ the Boltzmann's constant), $T_{\tau}$ is the imaginary-time
ordering operator,
$\langle \cdots \rangle$ is a thermal average, and $\gamma, \gamma' = (x,y,z)$ label
Cartesian components.
For our purposes, it is convenient to write the current operator in
(\ref{current-current-definition}) in terms of the Nambu representation of
the field
operators

\[ \Psi({\mathbf r})  =  \left( \begin{array}{c}
\psi_{\uparrow}({\mathbf r}) \\
                            \psi_{\downarrow}^{\dagger}({\mathbf r}) \end{array}
                            \right) \]

\noindent
as follows:
\begin{equation}
{\mathbf j}({\mathbf r})  =  \frac{1}{2im}  \sum_{l=1}^{2}  \left(
\Psi_{\ell}^{\dagger}({\mathbf r}) \nabla \Psi_{\ell}({\mathbf r})  - 
(\nabla \Psi_{\ell}^{\dagger}({\mathbf r})) \Psi_{\ell}({\mathbf r}) \right)
\label{current-definition}
\end{equation}
$m$ being the fermionic mass.

The current correlation function (\ref{current-current-definition}) can be
expressed in terms of the single-particle Green's function
\begin{equation}
{\mathcal G}(1,2)  =  -  \langle T_{\tau}[\Psi(1) \Psi^{\dagger}(2)] \rangle
\label{G-1}
\end{equation}
and of the two-particle Green's function
\begin{equation}
{\mathcal G}_{2}(1,2;1',2')  =  \langle T_{\tau}[\Psi(1) \Psi(2) 
\Psi^{\dagger}(2') \Psi^{\dagger}(1')] \rangle 
\label{G-2}
\end{equation}
(with the short-hand notation $1=({\mathbf r}_{1}, \tau_{1}, \ell_{1})$ and
so on, in terms of
the Nambu spinor components) as follows.
Let
\begin{equation}
L(1,2;1',2')  = {\mathcal G}_{2}(1,2;1',2') - {\mathcal G}(1,1')
{\mathcal G}(2,2')
\label{L}
\end{equation}
be the \emph{two-particle correlation function\/}, which satisfies the
Bethe-Salpeter equation
\begin{eqnarray}
L(1,2;1',2') & = & - {\mathcal G}(1,2') {\mathcal G}(2,1') +  \int \!d3456 
\; {\mathcal G}(1,3)
\nonumber   \\
& \times & {\mathcal G}(6,1') 
\Xi(3,5;6,4) L(4,2;5,2')
\label{Bethe-Salpeter}
\end{eqnarray}
where
\begin{equation}
\Xi(1,2;1',2')  =  \frac{\delta \Sigma(1,1')}{\delta {\mathcal G}(2',2)}
\label{2-p-int}
\end{equation}
is an \emph{effective two-particle interaction\/}.
Equation (\ref{Bethe-Salpeter}) can be formally solved in terms of the
\emph{many-particle
T-matrix\/}, defined as the solution to the equation \cite{Strinati-RNC}
\begin{eqnarray}
T(1,2;1',2') & = & \Xi(1,2;1',2')+\int \!d3456 \; \Xi(1,4;1',3)
\nonumber   \\
& \times & {\mathcal G}(3,6) 
{\mathcal G}(5,4)  T(6,2;5,2')
\label{many-p-T-matrix}
\end{eqnarray}
by writing
\begin{eqnarray}
-  L(1,2;1',2') & = & {\mathcal G}(1,2')  {\mathcal G}(2,1')+\int d3456 \; 
{\mathcal G}(1,3)
\nonumber   \\
& \times &  {\mathcal G}(6,1')  T(3,5;6,4) 
                          {\mathcal G}(4,2')  {\mathcal G}(2,5) .
\label{L-T}
\end{eqnarray}
In terms of the above quantities, one obtains the following expression for
the current
correlation function (\ref{current-current-definition}):
\begin{eqnarray}
\chi_{\gamma, \gamma'}({\mathbf r}\tau,{\mathbf r'}\tau') & = &
\frac{1}{(2m)^{2}}  \left(\nabla_{\gamma}  - 
\nabla^{''}_{\gamma}\right) 
\left(\nabla^{'}_{\gamma'}  -  \nabla^{'''}_{\gamma'}\right) 
\sum_{\ell,\ell'} \nonumber \\
& \times &
L({\mathbf r''}\tau\ell,{\mathbf r'''}\tau'\ell';{\mathbf r}\tau^{+}\ell,
\mathbf{r'}\tau'^{+}\ell')
\vert_{{\mathbf r''}={\mathbf r},{\mathbf r'''}={\mathbf r'}} .\nonumber \\
&&
\label{chi-L}
\end{eqnarray}
In Fourier space, the above relation reads (in three dimensions):
\begin{eqnarray}
\chi_{\gamma, \gamma'}({\mathbf q}, \Omega_{\nu}) & = &
\frac{1}{(2mi)^{2}}  \sum_{\ell,\ell'}  \int \!
\frac{d{\mathbf p}}{(2\pi)^{3}} 
 \int \! \frac{d{\mathbf p'}}{(2\pi)^{3}}  \frac{1}{\beta} \sum_{n}
e^{i\omega_{n}\eta} \nonumber \\
&\times &\frac{1}{\beta}\sum_{n'} e^{i\omega_{n'}\eta}
 (2p_{\gamma}+q_{\gamma})  (2p'_{\gamma'}+q_{\gamma'}) 
\nonumber \\
&\times & L^{\ell \ell'}_{\ell \ell'}
({\mathbf p}\omega_{n},{\mathbf p'}\omega_{n'};{\mathbf q}\Omega_{\nu})
\label{chi-L-FT}
\end{eqnarray}
where ${\mathbf q}$, ${\mathbf p}$, and ${\mathbf p'}$ are wave vectors,
$\Omega_{\nu}=2\pi\nu/\beta$
($\nu$ integer) is a bosonic Matsubara frequency, while
$\omega_{n}=(2n+1)\pi/\beta$ and
$\omega_{n}=(2n'+1)\pi/\beta$ ($n,n'$ integers) are fermionic Matsubara
frequencies ($\eta$ being
a positive infinitesimal).
The conventions for the (incoming and outgoing) four-vectors (with
$q=({\mathbf q},\Omega_{\nu})$,
$p=({\mathbf p},\omega_{n})$, and $p'=({\mathbf p'},\omega_{n'})$) and for
the Nambu  indices are
shown in Fig.~1, both for $L$ and the many-particle
$T$-matrix.\cite{footnote-Nambu-arrows}

\begin{figure}
\centerline{\epsfig{figure=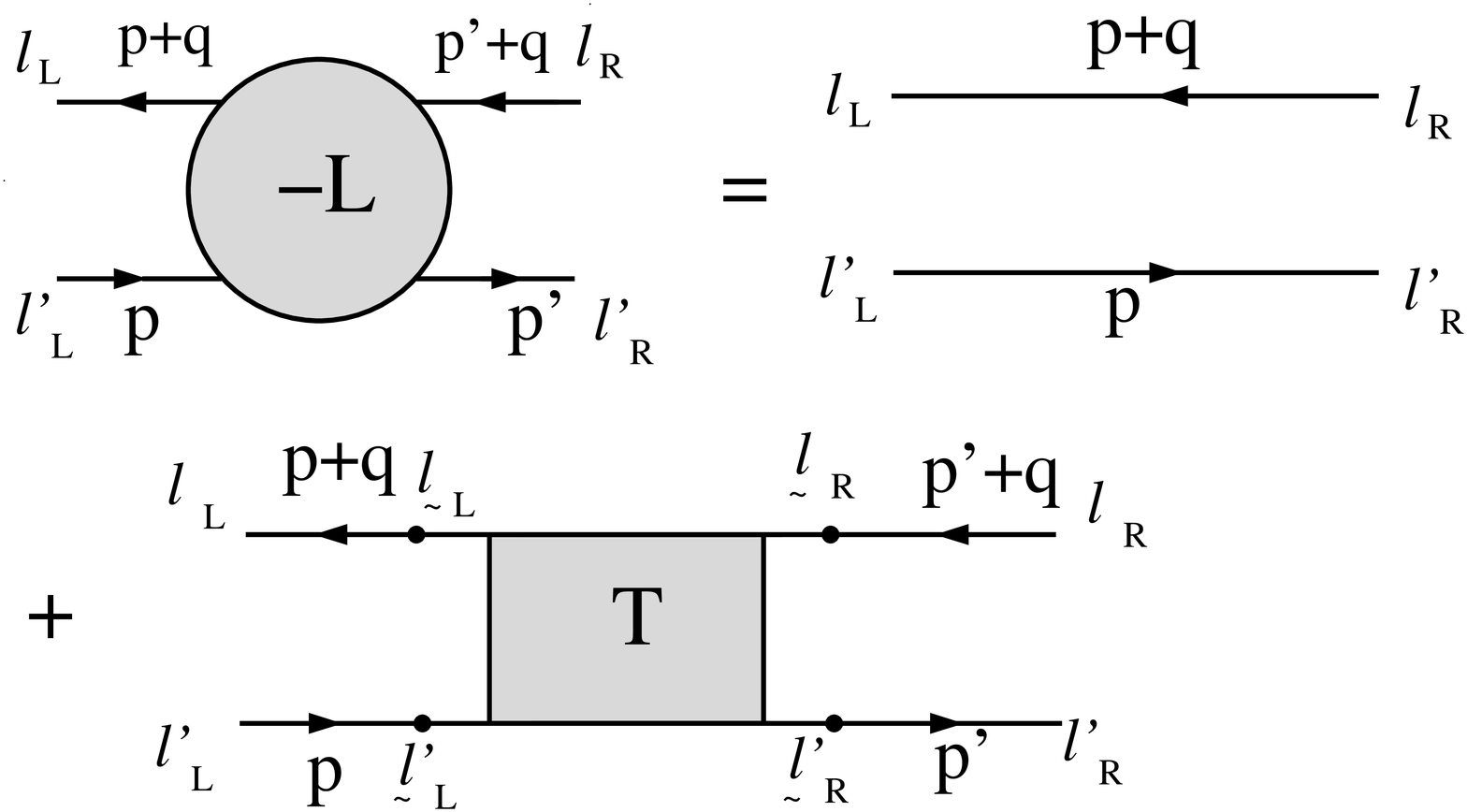,width=6.8cm}}
\vspace{0.1cm}
\caption{Graphical representation of the two-particle correlation
function ($L$) in
terms of the many-particle $T$-matrix ($T$) (four-momenta and Nambu
indices are indicated).}
\label{figure2}
\end{figure}

The above equations hold quite generally, regardless of the specific
approximation for the
kernel $\Xi$ defined in Eq.~(\ref{2-p-int}).
The BCS approximation is made manifest at this point by setting:
\begin{eqnarray}
\Xi_{BCS}(1,2;1',2') & = &
\frac{\delta \Sigma_{BCS}(1,1')}{\delta {\mathcal G}_{BCS}(2',2)}
\label{csibcs}   \\
& = & -  \tau^{3}_{\ell_{1}\ell_{2'}} \delta(x_{1}-x_{2'}) 
V(x_{1}^{+}-x_{1'}) \nonumber \\
& \times &\delta(x_{1'}-x_{2})  \tau^{3}_{\ell_{1'}\ell_{2}}  (1  - 
\delta_{\ell_{1}\ell_{1'}})
\nonumber
\end{eqnarray}
where, as usual, only the off-diagonal terms of the BCS self-energy have
been retained
($\tau^{3}$ being a Pauli matrix, and with the additional short-hand notation
$x_{1}=({\mathbf r}_{1}, \tau_{1})$ and so on).
Note that the factor $V(x_{1}^{+}-x_{1'})= \delta(\tau_{1}^{+}-\tau_{1'})
V({\mathbf r_{1}}-{\mathbf r_{1'}})$ represents the (instantaneous)
attractive effective fermionic interaction, as specified in Appendix A.

The first term on the right-hand side of Eq.~(\ref{L-T}),
when inserted in Eq.~(\ref{chi-L}), yields the standard BCS ``bubble'' 
contribution to the current correlation
function \cite{Schrieffer,FW}, \emph{provided\/} the BCS single-particle
Green's functions
\begin{eqnarray}
{\mathcal G}_{1 1}({\mathbf k},\omega_n)  & = &  - \frac{\xi({\mathbf k})
+ i \omega_n}
{E({\mathbf k})^2 + \omega_n^2}  =  -  {\mathcal G}_{2
2}({\mathbf -k},-\omega_n)  \nonumber  \\
{\mathcal G}_{2 1}({\mathbf k},\omega_n)  & = &  \frac{\Delta}
{E({\mathbf k})^2 + \omega_n^2}  =  {\mathcal G}_{1 2}({\mathbf k},\omega_n)
\label{BCS-Green-function}
\end{eqnarray}
are used. In the above expressions,
$\xi({\mathbf k})={\mathbf k}^{2}/(2m) - \mu$
($\mu$ being the chemical potential) and
$E({\mathbf k})=\sqrt{\xi({\mathbf k})^{2}+\Delta^{2}}$
for an isotropic ($s$-wave) gap function $\Delta$.
The BCS bubble contribution to the current correlation function taken alone
does not fulfill
the conservation laws (and, in particular, the \emph{longitudinal\/} f-sum
rule\cite{Schrieffer}); 
yet it gives a reasonable account of the \emph{transverse\/} current
correlation function in the
weak-coupling limit \cite{Schrieffer,FW}.

Quite generally, the \emph{superfluid density\/} $\rho_{s}$ can be obtained
from the knowledge
of the transverse component of the current correlation function as follows
~\cite{baymrhos}:
\begin{equation}
\sum_{\gamma,\gamma'}  \hat{t}_{\gamma}\, 
\chi_{\gamma,\gamma'} ({\mathbf q} \rightarrow 0,\Omega_{\nu} = 0) 
\, \hat{t}_{\gamma'}
 =  -  \frac{(n-\rho_{s})}{m}
\label{superfluid-density}
\end{equation}
where $n$ is the density and $\hat{t}$ is a unit vector transverse to
$\hat{q}={\mathbf q}/|{\mathbf q}|$.
In particular, inserting the BCS bubble contribution to the current 
correlation function in Eq.~(\ref{superfluid-density})
produces a meaningful overall temperature dependence of the superfluid
density in the weak-coupling limit,
not only for an $s$-wave \cite{FW} but also for a $d$-wave gap function
\cite{SD-d-wave-1,SD-d-wave-2}.
In the strong-coupling limit (and for temperatures small compared with the
dissociation energy
of the composite bosons), on the other hand, the BCS bubble contribution to
the current correlation
function vanishes, thus requiring consideration of additional diagrammatic
contributions.

Still within the BCS approximation, consideration of the second term on the 
right-hand side of Eq.~(\ref{L-T}) is required to fulfill conservation 
laws, provided that BCS single-particle Green's functions 
(\ref{BCS-Green-function}) are consistently used \cite{Schrieffer}.
In particular, it can be shown that the {\em longitudinal f-sum rule\/}~\cite{baymrhos}
\begin{equation}
\sum_{\gamma,\gamma'}  \hat{q}_{\gamma}\, 
\chi_{\gamma,\gamma'}({\mathbf q} \rightarrow 0,\Omega_{\nu} = 0)\, 
\hat{q}_{\gamma'}  = 
-  \frac{n}{m}                                                               
\label{SR}
\end{equation}
is fulfilled at any coupling in the static limit when this term is 
properly included.

It may further be of interest to obtain the full wave-vector and frequency
dependent 
current correlation function when the kernel $\Xi$ is taken within the BCS
approximation.
To this end, it is necessary to solve the integral equation
(\ref{many-p-T-matrix}) in Fourier space
with the BCS form (\ref{csibcs}) of the kernel, obtained from the
off-diagonal terms of the BCS
self-energy.
With this restriction, only four elements of the many particle $T$-matrix
with $\ell_{L}\neq\ell'_{L}$
and $\ell_{R}\neq\ell'_{R}$ survive (cf. Fig.~1) [we adopt the convention
$1 \leftrightarrow (\ell=1,\ell'=2)$
and $2 \leftrightarrow (\ell=2,\ell'=1)$ to label the nonvanishing matrix
elements of the $T$-matrix].
In addition, when the attractive effective fermionic interaction is taken
of the form of a
\emph{contact potential\/} with strength $v_{0}$ (cf. Appendix A), these
four independent elements
can be shown to satisfy the following algebraic equation:
\begin{eqnarray}
&&\left( \begin{array}{cc} T_{11}(q)  & T_{12}(q) \\ T_{21}(q) & T_{22}(q)
\end{array} \right)
=  v_{0} \left( \begin{array}{cc} 1 & 0 \\ 0 & 1 \end{array} \right)
\label{T-matrix-algebraic} \nonumber \\
&& + \, v_{0}  \left( \begin{array}{cc} -\Pi_{11}(q) & \Pi_{12}(q) \\
\Pi_{21}(q) & -\Pi_{22}(q) \end{array} \right)
\left( \begin{array}{cc} T_{11}(q) & T_{12}(q) \\ T_{21}(q) & T_{22}(q)
\end{array} \right)   .
\end{eqnarray}
In this expression,
\begin{equation}
\Pi_{11}(q)  =  \int \! \frac{d {\mathbf p}}{(2\pi)^{3}} 
\frac{1}{\beta}  \sum_{n} 
{\mathcal G}_{11}(p+q)  {\mathcal G}_{11}(-p)  =  \Pi_{22}(-q)
\label{Pi-11}
\end{equation}
and
\begin{equation}
\Pi_{12}(q)  =  \int \! \frac{d {\mathbf p}}{(2\pi)^{3}} 
\frac{1}{\beta}  \sum_{n} 
{\mathcal G}_{12}(p+q)  {\mathcal G}_{12}(-p)  =  \Pi_{21}(q)
\label{Pi-12}
\end{equation}
are the elementary rungs on which the many-particle $T$-matrix is built up
within the BCS approximation.

An equation similar to (\ref{T-matrix-algebraic}) holds for the
particle-particle ladder in the
broken-symmetry phase (cf. Appendix A).
Comparison of the two equations leads us to identify:
\begin{equation}
\left\{ \begin{array}{ll} T_{11}(q) = -\Gamma_{11}(q)  , & T_{12}(q)
= \Gamma_{12}(q) \\
T_{21}(q) = \Gamma_{21}(q)  , & T_{22}(q) = -\Gamma_{22}(q)  .
\end{array} \right.
\label{T-Gamma}
\end{equation}
In particular, the results obtained in Appendix A in the strong-coupling
limit (whereby
$\beta \mu \rightarrow -\infty$) can be exploited to write:
\begin{equation}
\left\{ \begin{array}{l} T_{11}(q)  =  (8 \pi / m^{2} a_{F}) 
{\mathcal G}'(q) \\
T_{12}(q)  =  (8 \pi / m^{2} a_{F})  {\mathcal G}_{21}'(q)
\end{array} \right.
\label{T-Bogoliubov}
\end{equation}
where $a_{F}$ is the fermionic scattering length (which is positive in the
strong-coupling limit)
and ${\mathcal G}'$ and ${\mathcal G}_{21}'$ are the normal and anomalous
noncondensate bosonic
Green's functions, respectively, in the Bogoliubov approximation \cite{FW}.

Within the above approximations, the many-particle $T$-matrix depends only
on $q$ and not
on all the incoming and outgoing four-vectors (cf. Fig.~1).
For this reason, the \emph{remainder\/} (i.e., other than the bubble)
contribution to the current
correlation function within the BCS approximation can be cast in the
following form:
\begin{eqnarray}
\chi^{rem}_{\gamma, \gamma'}(q) & = & I_{\gamma}(q)  T_{11}(q) 
I_{\gamma'}(q)  + 
                                      I_{\gamma}(q)  T_{12}(q) 
I_{\gamma'}(-q)   \nonumber \\
                                & + & I_{\gamma}(-q)  T_{21}(q) 
I_{\gamma'}(q)  + 
                                      I_{\gamma}(-q)  T_{22}(q) 
I_{\gamma'}(-q)  \nonumber \\
&&\label{factorized-form}
\end{eqnarray}
where we have set
\begin{equation}
I_{\gamma}(q)  =  \frac{1}{m}  \int \! \frac{d
{\mathbf p}}{(2\pi)^{3}}  \frac{1}{\beta}  \sum_{n} 
(2p_{\gamma}+q_{\gamma})  {\mathcal G}_{11}(p+q) {\mathcal G}_{21}(p)
 .            \label{definition-I}
\end{equation}
It can be readily shown that in the strong-coupling limit $I_{\gamma}(q)$
becomes
\begin{equation}
I_{\gamma}(q)  \simeq  -  q_{\gamma}  \frac{m a_{F} \Delta}{16 \pi}
 \, ,\label{approximation-I}
\end{equation}
so that the expression (\ref{factorized-form}) becomes eventually with the
help of Eq.~(\ref{T-Bogoliubov}):
\begin{equation}
\chi^{rem}_{\gamma, \gamma'}(q)  \simeq  q_{\gamma}  q_{\gamma'} 
n_{0} 
\frac{1}{(2m)^{2}}  \left[ {\mathcal G}'(q)  +  {\mathcal G}'(-q) -
  2 
{\mathcal G}'_{21}(q) \right]
\label{chi-rem-final}
\end{equation}
where the result (\ref{alpha-Delta}) (relating the general definition of
the \emph{condensate density\/} $n_{0} = |\alpha|^{2}$ to the off-diagonal 
single-particle BCS Green's function ${\mathcal G}_{12}$ in the strong-coupling
limit) has been exploited.
Note that, when the diagonal single-particle BCS Green's function 
${\mathcal G}_{11}$ is used to evaluate the particle density $n$, a relation 
similar to (\ref{alpha-Delta}) holds for $n/2$ in the place of
$n_{0}$, implying $n_{0} = n/2 $ at the level of the above
approximations.\cite{footnote-PPS}

The expression (\ref{chi-rem-final}) is purely
\emph{longitudinal\/}.
This expression can be thus compared with the longitudinal part of the
current correlation
function for a system of bosons within the Bogoliubov approximation (cf.
Appendix C).
Identifying $n/2$ with the bosonic density $n_{B}$ and $2m$ with the
bosonic mass $m_{B}$
(and neglecting the difference between the bosonic
density $n_{B}$  and the \emph{condensate density\/} $n_{0}$), one sees that 
the Bogoliubov result (\ref{chi-final}) is
four times smaller than the fermionic result (\ref{chi-rem-final}) in the
strong-coupling
limit.
This overall factor of four stems from having considered the 
current associated with the particle number and not with the charge density.
[This is equivalent to setting formally $e=1$ for the fermionic charge and
$e_{B}=2$ for the
bosonic charge.]
From Eq.~(\ref{chi-rem-final}) it can also be explicitly verified that the
f-sum rule
(\ref{SR}) is satisfied at this order of approximation.

Since the current correlation function for bosons within the Bogoliubov
approximation contains
not only a longitudinal but also a \emph{transverse\/} part (cf. Appendix
C), in the
strong-coupling limit the BCS approximation for the current correlation
function clearly fails
to recover the expression of the current correlation function within the
Bogoliubov approximation
for the composite bosons.
It is therefore necessary to search for additional fermionic diagrams for the
current correlation function in the broken-symmetry phase, which recover the 
transverse Bogoliubov result in the strong-coupling limit.
This task is achieved in the next Section.

\section{Transverse current correlation function in the broken-symmetry
phase from the analogy with the Bogoliubov result}

The simplest approximation for a Bose-condensed system is the so-called 
``Bogoliubov approximation'', that relies on the system being ``dilute''
\cite{Beliaev,Popov,FW}.
When one is interested in the crossover from a fermionic description of
physical quantities
in weak-coupling to a description in terms of (composite) bosons in
strong-coupling, it is natural
to require the (composite) bosons to be described by the Bogoliubov
approximation, at least as a first
significant step toward a more complete description of the strong-coupling
limit.

Concernig specifically the current correlation function, the form 
(\ref{chi-final})
for the  other-than-longitudinal contribution within the Bogoliubov
approximation suggests 
that the fermionic diagrams to be considered should contain two
particle-particle ladder
propagators (of the type discussed in Appendix A).
In analogy with a similar connection between fermionic and bosonic diagrams
established
in the normal phase \cite{SPL}, the Aslamazov-Larkin  \cite{AL} type
diagram of Fig.~2(a)
in the broken-symmetry phase appears to be a good candidate for our purposes.
The corresponding contribution to the current correlation function reads:
\begin{eqnarray}
\chi^{AL}_{\gamma, \gamma'}(Q) & = & -  \frac{1}{(2m)^{2}}  \int \!
\frac{d {\mathbf p}}{(2\pi)^{3}} 
\frac{1}{\beta}  \sum_{n}  \int \! \frac{d {\mathbf p'}}{(2\pi)^{3}} 
\frac{1}{\beta}  \sum_{n'} 
 \nonumber \\
&\times & \int \! \frac{d {\mathbf q}}{(2\pi)^{3}}  \frac{1}{\beta}  
\sum_{\nu} 
\sum_{ \ell_{L} \ell'_{L} \ell''_{L}}  \sum_{\ell_{R} \ell'_{R}
\ell''_{R}}    (2p_{\gamma} + 2q_{\gamma} + Q_{\gamma})      \nonumber \\
& \times &  (2p'_{\gamma'} + 2q_{\gamma'} + Q_{\gamma'}) \Gamma_{\ell'_{R}
\ell'_{L}}(q) \Gamma_{\ell''_{L}\ell''_{R}}(q+Q)\nonumber \\
& \times &{\mathcal G}_{\ell_{L} \ell'_{L}}(p+q)  {\mathcal G}_{\ell''_{L}
\ell_{L}}(p+q+Q) 
{\mathcal G}_{\ell''_{L} \ell'_{L}}(-p)
\nonumber \\
& \times & {\mathcal G}_{\ell'_{R} \ell_{R}}(p'+q)  {\mathcal G}_{\ell_{R}
\ell''_{R}}(p'+q+Q) 
{\mathcal G}_{\ell'_{R} \ell''_{R}}(-p') \nonumber\\
&&\label{AL-complete}
\end{eqnarray}
where $(p,p')$ and $(q,Q)$ are fermionic and bosonic four-vectors,
respectively, and the
minus sign is due to the presence of an additional fermion loop with
respect to the BCS
fermionic bubble already considered in Section II.
In the expression (\ref{AL-complete}) we have taken into account the
restrictions on the
Nambu  indices of the particle-particle ladder propagator, due to the
regularization of
the potential we have adopted.
Note that all single-particle Green's functions in Eq.~(\ref{AL-complete})
are meant to be
taken within the BCS approximation (\ref{BCS-Green-function}).

\begin{figure}
\centerline{\epsfig{figure=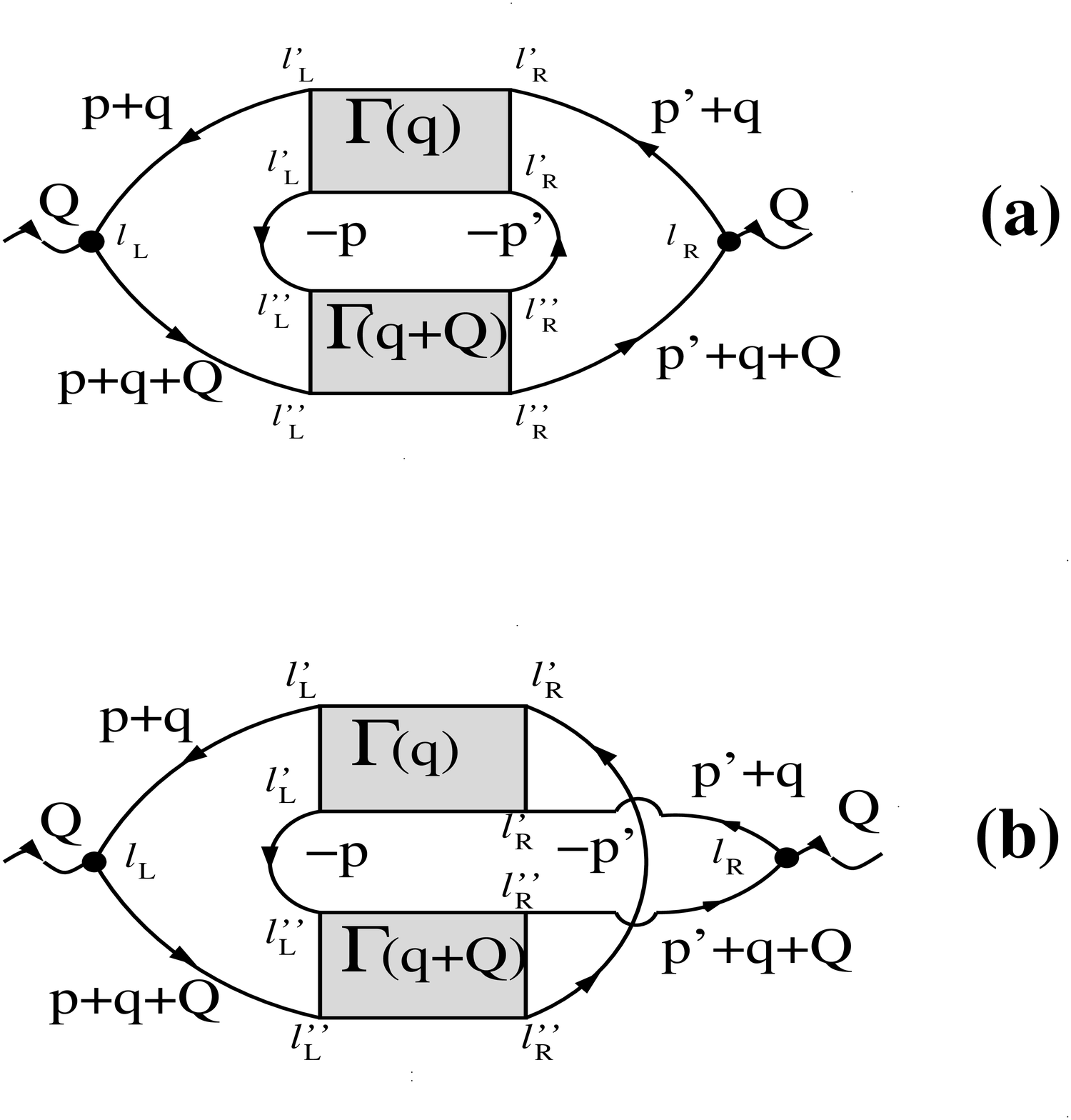,width=6.1cm}}
\caption{(a) AL type diagram for the current correlation function in
the broken-symmetry
phase; (b) Diagram topologically nonequivalent to (a) but with the same
value for a contact
potential.
}
\end{figure}

Although the expression (\ref{AL-complete}) formally holds for all
coupling, its calculation
gets considerably simplified \emph{in the strong-coupling limit\/} of
interest, whereby the
off-diagonal single-particle BCS Green's functions can be neglected with
respect to the diagonal ones.
In this case Eq.~(\ref{AL-complete}) reduces to:
\begin{eqnarray}
\chi^{AL}_{\gamma, \gamma'}(Q) & \simeq & -  \int \! \frac{d
{\mathbf q}}{(2\pi)^{3}} 
\frac{1}{\beta}  \sum_{\nu}  \sum_{\ell_{L} \ell_{R}} 
J_{\gamma}^{\ell_{L}}(q,Q)\,  \Gamma_{\ell_{R} \ell_{L}}(q) \nonumber \\
&\times &\Gamma_{\ell_{L} \ell_{R}}(q+Q)\,
 J_{\gamma'}^{\ell_{R}}(q,Q)\label{AL-strong-limit}
\end{eqnarray}
where we have introduced the vertex
\begin{eqnarray}
J_{\gamma}^{\ell}(q,Q) & = & \frac{1}{2m}  \int \! \frac{d
{\mathbf p}}{(2\pi)^{3}} 
\frac{1}{\beta}  \sum_{n}  (2p_{\gamma} + 2q_{\gamma} + Q_{\gamma}) \nonumber \\
&\times&{\mathcal G}_{\ell \ell}(p+q)
 {\mathcal G}_{\ell \ell}(p+q+Q)  {\mathcal G}_{\ell \ell}(-p) \,  .
\label{vertex-AL}
\end{eqnarray}
Since $J_{\gamma}^{2}(q,Q)  = J_{\gamma}^{1}(-q,-Q)$ owing to the
symmetry relations in Eq.~(\ref{BCS-Green-function}), it is sufficient to evaluate $J_{\gamma}^{1}(q,Q)$.
One obtains:
\begin{equation}
J_{\gamma}^{1}(q,Q)  \simeq  -  \frac{(2 q_{\gamma} + Q_{\gamma})}{2m} 
\frac{m^{2}  a_{F}}{16  \pi}
\label{vertex-AL-approximate}
\end{equation}
in agreement with the value obtained for the corresponding vertex of the AL
diagram in
the normal phase \cite{SPL}.
Making explicit the Nambu  indices in Eq.~(\ref{AL-strong-limit}) and 
recalling Eqs.~(\ref{T-Gamma}) and (\ref{T-Bogoliubov}) for the connection 
between the elements of the particle-particle ladder and the 
bosonic propagators
within the
Bogoliubov approximation, we obtain eventually in the strong-coupling limit:
\begin{eqnarray}
\chi^{AL}_{\gamma, \gamma'}(Q) & \simeq & -  \frac{1}{(2m)^{2}} 
\frac{1}{4} 
\int \! \frac{d {\mathbf q}}{(2\pi)^{3}}  \frac{1}{\beta}  \sum_{\nu} 
(2 q_{\gamma} + Q_{\gamma}) 
\nonumber \\
& \times & (2 q_{\gamma'} + Q_{\gamma'})\left[ {\mathcal G}'(q)  {\mathcal G}'(q+Q) + {\mathcal G}'(-q)\nonumber \right.\\
&\times & \left.{\mathcal G}'(-q-Q)  -  2  {\mathcal G}_{21}'(q) 
{\mathcal G}_{21}'(q+Q) \right]   .
\label{AL-Bogoliubov}
\end{eqnarray}
This expression has to be compared with the current correlation function
for bosons
within the Bogoliubov approximation, i.e., with the second term on
the right-hand
side of Eq.~(\ref{chi-final}).
The comparison shows that expression (\ref{AL-Bogoliubov}) is only two
times larger than
its Bogoliubov counterpart, while one would expect it to be four times
larger according
to the discussion following Eq.~(\ref{chi-rem-final}).
The missing factor of two is provided by an additional diagram [cf. Fig.~2(b)],
which is topologically nonequivalent to the diagram of Fig.~2(a) but has the 
same value for a contact potential.

In the weak-coupling limit, on the other hand, the AL  type
diagrams of Fig.~2 in the
broken-symmetry phase represent fluctuation contributions to the
ordinary BCS bubble approximation, and merge with the usual Aslamazov-Larkin 
contribution~\cite{AL}
when approaching the critical temperature from below.

Having succeeded in obtaining the Bogoliubov result for the transverse part
of the current correlation function in the broken-symmetry phase as the 
strong-coupling limit of the fermionic diagrams
of Fig.~2, we pass now to determine the evolution
of the transverse
part of the current correlation function from weak to strong coupling, by
approximating
the current correlation function in all coupling regimes as the sum of the
standard BCS
bubble diagram and the diagrams of Fig.~2.
In particular, we shall be interested in the \emph{static limit\/} of these
diagrams,
from which the superfluid density $\rho_{s}$ can be extracted as
${\mathbf q} \rightarrow 0$
according to Eq.~(\ref{superfluid-density}) and the Pippard kernel can be
obtained by retaining
finite values of ${\mathbf q}$.
The numerical results obtained, from weak to strong
coupling, for the
superfluid density as a function of temperature and for the Pippard kernel
at zero temperature are discussed in the next Section.
\section{Numerical results for the superfluid density and the Pippard-like
kernel from weak
to strong coupling}

In this Section, we evaluate numerically the AL type contribution
(\ref{AL-complete}) to the
current correlation function for vanishing external frequency from weak to
strong coupling,
in addition to the standard BCS bubble contribution.
From this calculation we extract the temperature dependence of the
superfluid density (when the external wave vector is also allowed to vanish) 
as well as the form of the Pippard-like kernel at zero temperature (by 
keeping the external wave vector finite).

While the numerical calculation of the BCS bubble is rather
straightforward, calculation of the expression
(\ref{AL-complete}) requires a non-negligible numerical effort.
Specifically, evaluation of the particle-particle ladder $\Gamma$ [cf.
Eqs.~(\ref{Gamma-solution})
and (\ref{A-definition})], where the frequency sum can be done
analytically, requires a
two-dimensional numerical integration; in addition, the four independent
vertices with three
single-particle Green's functions (of which the expression
(\ref{vertex-AL}) represents a degenerate
case), where again the frequency sum can be done analytically, require a
two-dimensional
(three-dimensional) numerical integration for vanishing (finite) external
wave vector; finally,
both the frequency sum and the wave vector integration over the bosonic
variable $q$ of the
$\Gamma$ functions have to be done numerically.
Special care has to be exerted when performing this frequency sum,
which requires explicit
inclusion of about 300 to 500 (positive) Matsubara frequencies, as well as a
careful treatment of the
tail of the frequency sum which we approximate by an integral extending up
to infinity.
The resulting function (to be eventually integrated over
the wave vector ${\mathbf q}$) is, in fact, affected by spurious oscillations 
for large values of  $|{\mathbf q}|$, which result
from incomplete cancellations of large numbers when the tail of the above
frequency sum is added
to the contribution of the finite set of Matsubara frequencies.
In addition, these oscillations are strongly amplified by the factor
${\mathbf q}^{4}$ originating
from the measure of the (three-dimensional) integral and from the
wave-vector dependence of the
vertices with three Green's functions.
As a consequence, we have pragmatically truncated the integral over
$|{\mathbf q}|$ where these
spurious oscillations start to appear.
We have consistently verified that this truncation procedure yields correct
results in the two
cases where analytic controls can be made, namely, in the
strong-coupling limit at low enough
temperatures and in the weak-coupling limit close to the (BCS) critical
temperature.

A comment on the procedure we have adopted to generate the values of the
gap function $\Delta$ and
chemical potential $\mu$ is in order at this point (these parameters
enter the single-particle
Green's functions (\ref{BCS-Green-function}) and, through them, all other
relevant quantities).
According to a standard procedure when following the evolution from BCS
superconductivity to BE condensation \cite{Eagles,Leggett}, 
$\Delta$ and $\mu$ are obtained by solving the coupled equations for 
the mean-field gap and particle number, in terms of the
Green's functions (\ref{BCS-Green-function}).
Strictly speaking, these equations are valid at low enough temperature
(compared with the superconducting critical temperature), and corrections to 
these equations should accordingly be considered at higher temperatures.
In this paper, we restrict to low enough temperatures, so that inclusion of
these corrections is not expected to be important.
Improved results at higher temperatures may then be obtained by supplying
the present calculation for the current correlation function with improved 
values of $\Delta$ and $\mu$, which take into account the effects of 
superconducting fluctuations below $T_{c}$.~\cite{footnote-PPS} 

\subsection{Superfluid density}

Once the AL type and the fermionic bubble contributions to the current
correlation function are
calculated as described above, the temperature dependence of the superfluid
density $\rho_{s}$
can be obtained via the definition (\ref{superfluid-density}) by setting
$\Omega_{\nu}=0$ and letting ${\mathbf q} \rightarrow 0$.

The low-temperature behavior of the fermionic bubble leads to the
well-known exponential suppression of the
\emph{normal\/} density $n-\rho_{s}$ when approaching zero temperature;
this exponential behavior is
governed by the BCS gap in weak coupling and by the magnitude of the
chemical potential in strong coupling.
These energy scales, in turn, limit the respective temperature range for
the validity of the exponential
behavior.
The low-temperature behavior of the AL type diagram in the strong-coupling
limit, on the other hand, yields
a $T^{4}$ increase of the normal density for increasing temperature,
in agreement with the Bogoliubov-Landau
expression \cite{FW} (cf. Appendix C).
In this limit, the $T^{4}$ behavior is valid for temperatures small
compared with the bosonic chemical
potential and is due to the \emph{linear\/} dispersion of the Bogoliubov
mode for small enough wave
vectors.
Since the poles of the particle-particle ladder $\Gamma$ yield a linear
dispersion even at intermediate
and weak coupling \cite{MPS} (giving origin to the so-called
Bogoliubov-Anderson mode \cite{Micnas-92}),
one expects the $T^{4}$ behavior from the AL type contribution to survive
at low enough temperature for
all coupling ranges.
This $T^{4}$ behavior is thus expected to dominate over the BCS exponential
behavior at low enough temperatures.
[Note, however, that the $T^{4}$ behavior is bound to cross over into a
different power-law behavior at
a finite characteristic temperature, as discussed below.]

\begin{figure}
\centerline{\epsfig{figure=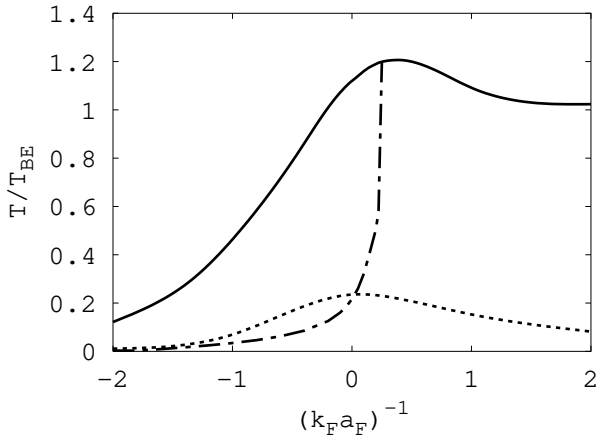,width=6.0cm,angle=-90}}
\vspace{0.1cm}
\caption{Temperatures below which (dashed-dotted line) the AL
type contribution to
the normal density is dominant with respect to the BCS contribution and below 
which (dotted line) the $T^{4}$ behavior of the AL type contribution sets in, 
versus the coupling parameter $(k_{F}a_{F})^{-1}$.
The superconducting critical temperature $T_{c}$ (full line) is also shown
for comparison.
[All temperatures are in units of the Bose-Einstein condensation
temperature $T_{BE}= 3.31 n_{B}^{2/3}/m_{B}$.]}
\end{figure}

An interesting question that can be answered with the present calculation is 
to determine the temperature range below which
the AL type contribution dominates over the BCS contribution on the
\emph{whole\/} coupling range, from weak to strong.
To this end, we have calculated the temperature at which these two
contributions equal each other for several values
of the coupling parameter $(k_{F}a_{F})^{-1}$ (where $k_{F}$ is the Fermi
wave vector and $a_{F}$ is the scattering length - cf. Appendix A).
[In practice, the whole crossover region can be spanned by varying
$(k_{F}a_{F})^{-1}$ about from $-1$ to $+1$; deviations from
 weak-coupling approximations when $(k_{F}a_{F})^{-1}\lesssim - 1$
and from strong-coupling approximations when $(k_{F}a_{F})^{-1}\gtrsim 1$
become, in fact, negligible for all practical purposes.]
In Fig.~3 this temperature is plotted versus $(k_{F}a_{F})^{-1}$, together
with the superconducting critical
temperature $T_{c}$ obtained following Ref.~\onlinecite{NSR} (cf.~also
Refs.~\onlinecite{Haussmann-2} and \onlinecite{PPSC}).
In the same plot, we have also reported the temperature at which the AL type 
contribution changes from the $T^{4}$ behavior to a different power law, 
from which one notes that the range of validity of the ``low-temperature''
behavior is significantly enhanced in the crossover region.
[It is, however, clear from the previous discussion that the results for
these two boundary temperatures are only indicative when the two 
temperatures become of order $T_{c}$.]
Note also that the region where the AL type contribution dominates over the
BCS bubble is resticted to very low temperatures in weak coupling, while it 
increases rapidly in the crossover region.
In particular, it reaches about one-fifth of the critical temperature
at $(k_{F}a_{F})^{-1}=0$, a coupling value for
which the chemical potential is still positive and remnants of the Fermi
surface survive (the chemical potential
changing sign at $(k_{F}a_{F})^{-1}=0.55$).
The inclusion of the AL type diagram turns thus out to be appreciable as
soon as one leaves the \emph{extreme\/}
weak-coupling region.
In the strong-coupling side of the crossover region (i.e., for
$(k_{F}a_{F})^{-1} \gtrsim 0.3$), the AL type
contribution definitely dominates over the BCS contribution for all
relevant temperatures.

\begin{figure}
\centerline{\epsfig{figure=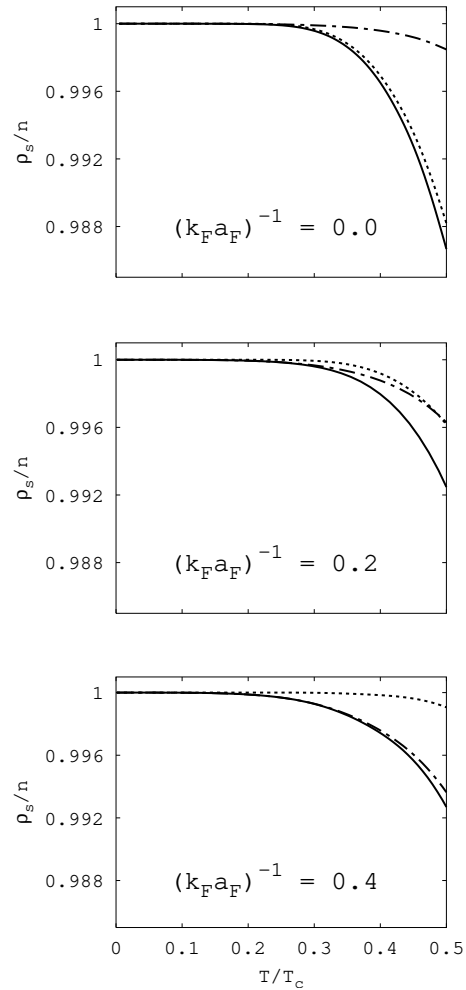,width=2.5in}}\
\vspace{0.2truecm}
\caption{Temperature dependence of the AL type (dashed-dotted line) and
BCS (dotted line) contributions
to the superfluid density, and sum of the two contributions (full
line), vs temperature when
 $(k_{F}a_{F})^{-1} = 0.0, 0.2, 0.4$, from top to bottom.
[The temperature is in units of $T_{c}$ as obtained from Fig.~3.]}
\end{figure}

The temperature dependence of the AL type and BCS contributions to the
superfluid density is shown
in Fig.~4 for three characteristic couplings in the crossover region,
namely, $(k_{F}a_{F})^{-1} = 0.0, 0.2, 0.4$.
We have extended this calculation up to halfway the critical temperature,
where the values of $\Delta$ and $\mu$ obtained as described above are still 
reliable.
Note that for $(k_{F}a_{F})^{-1} = 0.0$ the BCS contribution
dominates over the AL type contribution, while
for $(k_{F}a_{F})^{-1} = 0.4$ the opposite occurs.
This implies that, even for the superfluid density, the crossover from
weak- to strong-coupling behavior occurs in a rather \emph{narrow\/} region 
of the coupling parameter $(k_{F}a_{F})^{-1}$.
The mechanism for this rapid crossover is that the BCS contribution becomes 
quickly negligible when $(k_{F}a_{F})^{-1}$ approaches zero from negative 
values ($(k_{F}a_{F})^{-1} = 0$ corresponds to the threshold for the occurrence
of a bound state in the associated two-body problem); the AL type
contribution, on the other hand, becomes quickly negligible when 
$(k_{F}a_{F})^{-1}$ approaches zero from positive values.
It turns thus out that there exists a finite interval of the parameter
$(k_{F}a_{F})^{-1}$ where \emph{both\/} contributions are small with respect 
to the full density $n$ (at fixed temperature).
From a physical point of view, one intepretes this result by noting that there
are two possible mechanisms to destroy the superfluid density,
namely, the pair-breaking effect on the weak-coupling side and the
excitation of collective modes on the strong-coupling
side.
As both mechanisms tend to be suppressed in the intermediate-coupling region,
the value of the superfluid density is correspondingly enhanced and 
superfluidity gets more robust against thermal fluctuations.

To establish comparison with the experimental data on
cuprate superconductors, the calculation
of the temperature dependence of the superfluid density should
realistically include also lattice effects and the $d$-wave character of the
gap function.
Nevertheless, we expect the above physical argument (about the
increasing robustness of the superfluid density
against thermal fluctuations when approaching the intermediate-coupling
region from both sides) to remain valid even in the presence of a $d$-wave 
gap function.
In fact, in the presence of a $d$-wave gap function the BCS contribution
gets suppressed even faster for increasing
coupling as soon as the chemical potential crosses the bottom of the
single-particle energy band, passing from a
linear-$T$ behavior \cite{SD-d-wave-1,SD-d-wave-2} to an ($s$-wave like)
exponential behavior; the AL type contribution
in the presence of a $d$-wave gap, on the other hand, should not be
appreciably modified with respect to the results
obtained with an $s$-wave gap, at least when the chemical potential remains
below the bottom of the single-particle band.

An example of a superconductor with an $s$-wave gap function, for which the
effective coupling may be sufficiently
strong that the effect of collective modes on the superfluid density could
be revealed, is the recently discovered $MgB_{2}$.
Experimental data for the temperature dependence of the penetration depth
(and thus of the superfluid density)
show a quadratic temperature dependence over most of the temperature range
up to $T_{c}$, but for noticeable
deviations in the low-temperature region which appear instead to follow a
quartic behavior \cite{MgB2}.
While the quadratic temperature dependence away from the low-temperature
region may simply result from an ordinary ($s$-wave) BCS contribution (which 
can indeed be rather well fitted
to a quadratic behavior upon
decreasing the temperature from $T_{c}$), the quartic behavior at low
temperature may instead be due to the presence of
collective modes.
[The occurrence of the above quadratic temperature dependence at high
enough temperatures has alternatively be considered as
an indication of the presence of nodes in the superconducting energy
gap of $MgB_{2}$.~\cite{MgB2}]
In Fig.~5 we report the superfluid density versus $T^{2}$, as obtained by
our calculations with both BCS and AL
type contributions included, for a specially chosen value of the coupling
($(k_{F}a_{F})^{-1} = -0.4$) such that
both the $T^{4}$ behavior (due to the AL type contribution at low
temperatures) \emph{and\/} the $T^{2}$ behavior
(due to the BCS contribution at higher temperatures) are evident.
Note that, to evidence the $T^{2}$ behavior at high enough temperatures,
the numerical results reported in Fig.~5
have been extended beyond their strict range of validity, where the values
of $\Delta$ and $\mu$ should rather be  
obtained by an improved calculation as mentioned above.
For this reason, the temperature scale in Fig.~5 has been normalized with
respect to the mean-field critical temperature $T_{{\rm MF}}$ at which 
$\Delta$ vanishes in the present calculation
(and not with respect to $T_{c}$ reported in Fig.~3).
Note also that, while the experimental curve reported in Ref.~\onlinecite{MgB2}
shows a change of convexity when passing from
the $T^{2}$ to the $T^{4}$ behavior, the theoretical curve shown in Fig.~5
mantains the same convexity over the whole
temperature range.

\begin{figure}
\centerline{\epsfig{figure=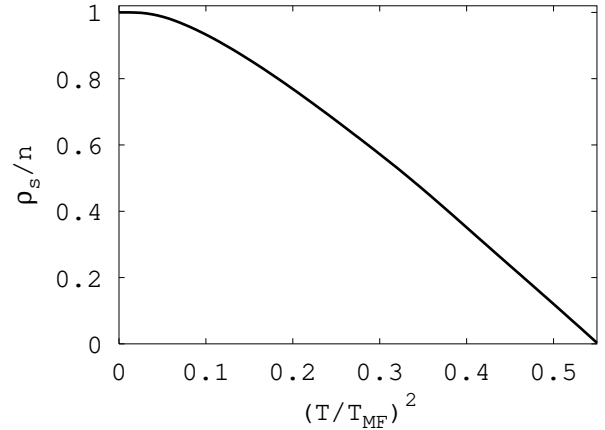,width=6.0cm,angle=-90}}
\vspace{0.2truecm}
\caption{Superfluid density vs $T^{2}$ when $(k_{F}a_{F})^{-1} =
-0.4$, including both
BCS and AL type contributions.
[The temperature is in units of the mean-field critical temperature.]}
\vspace{0.1cm}
\end{figure}

Care should definitely be exerted when trying to compare the temperature
dependence of the superfluid density,
obtained also with the inclusion of collective modes, with the experimental
data for superconductors
where density fluctuations produce charge unbalance.
In this case, the Bogoliubov mode with a linear dispersion at small wave
vectors should turn into a gapped \emph{plasmon\/}
mode owing to the long-range Coulomb interaction.
Dissipation effects have also been recently shown to play an important role
in restoring a power-law behavior for the
superfluid density at low temperature, albeit with a different exponent
from that obtained by considering only neutral
collective modes \cite{plasmon}.
Renormalization of the chemical potential (which drives the system from the
BCS to the BE limit) has,
however, not been considered in this context.
A more realistic approach to the calculation of the low-temperature
behavior of the superfluid density would then
include both Coulomb and damping effects on top of the treatment presented
in this paper, which is appropriate to follow the crossover from the BCS to 
the BE limit.
Assessment of the relative importance of quasi-particle excitations and
collective modes to the superfluid
density, in fact, can be realistically done only in the context of this
crossover when the chemical potential is
allowed to be properly renormalized.
This more complete approach is deferred to future work.

Notwithstanding these limitations, it is interesting to exploit further our
calculation for the current correlation
at zero frequency, to obtain the form of the Pippard-like kernel from weak
to strong coupling at zero temperature.

\subsection{Pippard-like kernel}

As discussed in the Introduction, two different length scales 
$\xi_{{\rm phase}}$
(describing correlation among different Cooper pairs or composite bosons) 
and $\xi_{{\rm pair}}$ (describing correlation between fermions in a pair) 
enter the BCS-BE crossover problem.
In particular, while $\xi_{{\rm phase}}\sim\xi_{{\rm pair}}$ in the
weak-coupling limit, $\xi_{{\rm phase}}\gg\xi_{{\rm pair}}$ in the 
strong-coupling limit.~\cite{PS-96} In addition, in the weak-coupling limit 
$\xi_{{\rm pair}}$ coincides with the Pippard coherence length $\xi_0$
characterizing the non-local relation between the supercurrent $\bf{j}$ and 
the vector potential $\bf{A}$.\cite{Pippard} It is then natural to ask which 
one of the
two lengths ($\xi_{{\rm pair}}$ or $\xi_{{\rm phase}}$) enters this relation 
between the 
supercurrent and the vector potential also in the intermediate- and 
strong-coupling regions. In this subsection, we consider the contribution to 
the 
${\bf j}$ vs ${\bf A}$ relation from the standard BCS bubble and the AL type 
diagram introduced in Section III, and show that $\xi_{{\rm pair}}$ and 
$\xi_{{\rm phase}}$ are associated with the two contributions, in the order.

Quite generally, to extract the Pippard-like kernel (whose precise definition 
will be given below) from the expression of the current correlation function 
in a form appropriate to numerical calculations,
it is convenient to identify from Eq.~(\ref{AL-complete}) the function
$G^{AL}({\mathbf q},{\mathbf p})$, such that 
\begin{eqnarray}
\chi^{AL}_{\gamma, \gamma'}({\mathbf Q},\Omega_{\nu}=0) & = & -  \frac{1}{(2m)^{2}}  \int \!
\frac{d {\mathbf q}}{(2\pi)^{3}} 
(2q_{\gamma} + Q_{\gamma})   \nonumber \\
& \times &(2q_{\gamma'} + Q_{\gamma'})\frac{G^{AL}({\mathbf q},{\mathbf q}+{\mathbf Q})}{2  |{\mathbf q}| 
|{\mathbf q}+{\mathbf Q}|}  \label{chi-G-AL}
\end{eqnarray}
for the transverse components (we shall restrict, in particular, to the 
zero-temperature limit).
With this definition, the function $G^{AL}({\mathbf q},{\mathbf p})$ reduces
to the expression (\ref{G}) of Appendix C
in the strong-coupling limit.

For all couplings, it turns out that
$G^{AL}({\mathbf q},{\mathbf p})=G^{AL}(|{\mathbf q}|,|{\mathbf p}|)$ and that
$G^{AL}(q,q)=0$, as in the Bogoliubov case (cf. Appendix C). [We have
verified these properties numerically for the AL case within a 5 \% accuracy.]
In this way, Eqs.~(\ref{approx-derivatives})-(\ref{g-integral}) of
Appendix C remain formally valid for all
couplings (with the label AL attached to the quantities $G(p,q)$ and $g(R)$
therein).

Strictly speaking, however, it is no longer possible to extract the 
delta-function from the AL contribution at arbitrary coupling 
as in Eq.~(\ref{g-decomposition}), since for composite bosons
$\lim_{p\to\infty}p\, G^{AL}(q,p)=0$ (instead of the 
behavior (\ref{f-definition}) for point-like bosons). For composite bosons the
coefficient ${\mathcal I}$ of Eq.~(\ref{g-decomposition}) vanishes and only 
the nonlocal 
contribution $g_n^{AL}(R)$ survives. Nonetheless, upon approaching the extreme 
strong-coupling limit (where the composite nature of the bosons no longer 
matters) the small-$R$ behavior of $g_n^{AL}(R)$  can be {\em effectively} 
assimilated to a delta-function (since its range extends over a 
length scale much smaller than the scale $\xi_{{\rm phase}}$ associated with
the remaining part of $g_n^{AL}(R)$), such that the expected behavior
for point-like bosons is properly recovered in this limit, as also shown below 
by the numerical results.

Similarly, the BCS bubble contribution can be calculated at arbitrary 
couplings according to the procedure of Appendix D, instead of relying on the 
standard BCS analytic calculation for weak-coupling\cite{FW}. In this case, 
one can identify the 
function $g_n^{{\rm BCS}}(R)$ for $R\neq 0$ and its integral 
${\mathcal I}^{{\rm BCS}}=-8\pi^3 m n$ (cf. Eq.~(\ref{I-valueapp})).
What happens to the BCS bubble contribution is that, in the weak-coupling 
limit, the function  $g_n^{{\rm BCS}}(R)$ contains a short-range part 
extending over the
scale $k_F^{-1}\ll\xi_{{\rm pair}}$, whose integral (from $R=0$ to 
$R=k_F^{-1}$) together with the delta-function contribution proportional to 
${\mathcal I}^{{\rm BCS}}$ cancels the diamagnetic contribution to the 
${\bf j}$ vs ${\bf A}$ relation, making only the nonlocal part of 
$g_n^{{\rm BCS}}(R)$ effectively surviving.

Putting together the results for the AL and BCS contributions, the desired 
relation between the induced current ${\mathbf j}$ and the vector potential 
${\mathbf A}$ acquires the form:
\begin{eqnarray}
j_{\gamma}({\mathbf r}) & = & -  \frac{n}{3 m  c}  A_{\gamma}({\mathbf r})
\nonumber \\
& + & \frac{1}{(2\pi)^{4}  m^{2}  c}  \int \! d{\mathbf r'} 
\frac{ R_{\gamma}  \sum_{\gamma'} R_{\gamma'} A_{\gamma'}({\mathbf r'})
}{R^{4}} \nonumber \\ 
&\times& \left( g^{BCS}_{n}(R)+  g^{AL}_{n}(R) \right)
\label{j-A-general-BCS-AL}
\end{eqnarray}
where ${\mathbf R}={\mathbf r}-{\mathbf r'}$ and $c$ is the light velocity.
This relation defines the \emph{Pippard-like kernel\/} at arbitrary couplings,
which is the sum of a local (London-like) and of a
nonlocal component.
Note that the local part of Eq.~(\ref{j-A-general-BCS-AL}) coincides with the 
local part of Eq.~(\ref{j-A-general-BCS-ALapp}) since the AL contribution 
lacks the delta function.

The results of the numerical calculation for the two (BCS and AL) 
contributions to Eq.~(\ref{j-A-general-BCS-AL}) 
show that, in the weak-coupling limit, the BCS 
contribution dominates over the AL counterpart even though both contributions 
extend over the same spatial range since 
$\xi_{{\rm pair}}\sim\xi_{{\rm phase}}$ in this limit. In the strong-coupling 
limit, on the other hand, the BCS bubble no longer contributes significantly 
to the response function since the nonlocal part 
$g^{BCS}_{n}(R)$ cancels the delta-function contribution over a scale
$\xi_{{\rm pair}}$ much smaller than the scale $\xi_{{\rm phase}}$ of the AL 
contribution. At arbitrary coupling, both BCS and AL contributions to 
Eq.~(\ref{j-A-general-BCS-AL}) need be explicitly considered.

\begin{figure}
\centerline{\epsfig{figure=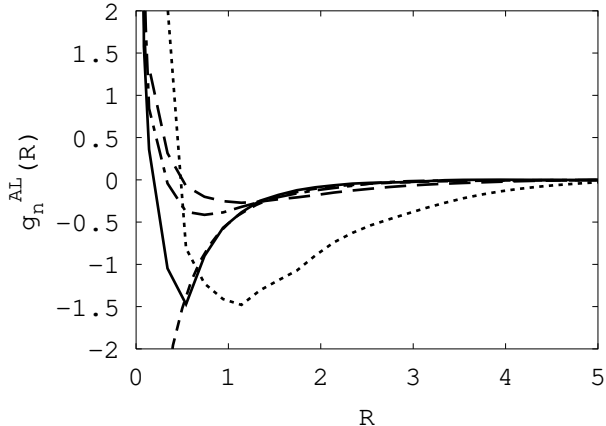,width=6.0cm,angle=-90}}
\caption{Function
 $g^{AL}_{n}(R)$ (in units of $m_B/(4 \xi_{\rm{phase}}^4)$) vs 
$R$ (in units of $\sqrt{2}\xi_{\rm{phase}}$)
for four different values of the coupling
parameter $(k_{F}a_{F})^{-1}$:
$-1.0$ (dotted line), $0.0$ (long-dashed line), $+1.0$ (dashed-dotted line),
$+2.0$ (full line). The purely bosonic $g_n^B(R)$ (short-dashed line) is also
shown for comparison. }
\end{figure}

In Fig.~6 the functions $g^{AL}_{n}(\tilde{R})$ is reported vs 
$\tilde{R}=R/(\sqrt{2}\xi_{{\rm phase}})$ for different values of the coupling 
parameter $(k_F a_F)^{-1}$. 
The relevant range of $g^{AL}_{n}(\tilde{R})$ extracted from this plot is
determined by the coherence length $\xi_{{\rm phase}}$ introduced in 
Ref.~\onlinecite{PS-96} to account for the
spatial correlations of the amplitude of
the order parameter (direct calculation\cite{MPS} of $k_{F} \xi_{{\rm phase}}$ 
yields the values $(1.7,0.7,0.8,1.1)$
corresponding to the values $(k_{F} a_{F})^{-1} = (-1.0,0.0,+1.0,+2.0)$
here considered).
[We have also verified that $k_{F} \xi_{{\rm phase}}$ extracted from the 
behavior 
of $g^{AL}_{n}(R)$  attains its minimum at $(k_{F}a_{F})^{-1} = +0.4$,
in agreement with the results from Ref.~\onlinecite{PS-96}.]
Note that the delta-function contribution to $g(R)$, which is present in the 
bosonic expression of Appendix C, is recovered in Fig.~6 by the 
progressive shrinking of the short range part of $g_n^{AL}(R)$ as the coupling
is increased; at the same time, the long-range part of $g_n^{AL}(R)$ in
Fig.~6 approaches the asymptotic behavior of the bosonic $g^{B}_n(R)$. We should warn, however, that extracting the short-range part of 
$g_n^{AL}(R)$ 
from numerical calculation of the AL type diagram is affected by sizable
numerical errors, since it requires an accurate treatment of the large 
wave-vector behavior of $G^{AL}(q,p)$ (this procedure, in turn, requires 
subtraction of large terms which should add up to zero).

\begin{figure}
\centerline{\epsfig{figure=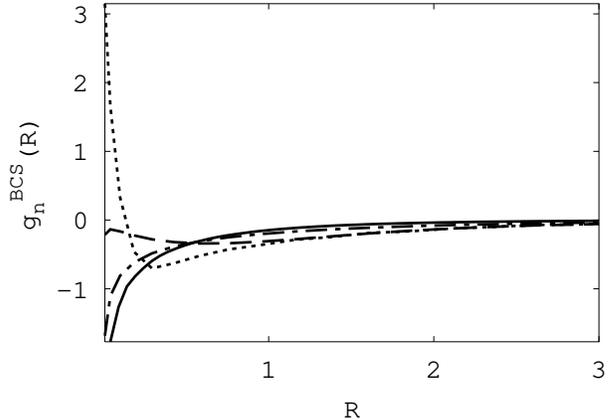,width=6.0cm,angle=-90}}
\caption{Function $g_n^{BCS}(R)$ (in units of $m k_F^4$) vs 
$R$ (in units of $\xi_{{\rm pair}}$) for four different values of the coupling
parameter $(k_{F}a_{F})^{-1}$:
$-2.0$ (dotted line), $-1.0$ (dashed line), $0.0$ (dashed-dotted line),
$+1.0$ (full line).} 
\end{figure}

In Fig.~7 the function $g_n^{BCS}(\tilde{R})$ is reported vs
 $\tilde{R}=R/\xi_{{\rm pair}}$ for different values of the coupling parameter
$(k_F a_F)^{-1}$.
The relevant range of $g^{BCS}_{n}(\tilde{R})$ is 
determined by the coherence length $\xi_{{\rm pair}}$
for two-fermion correlation \cite{PS-94,PS-96}, which is monotonically
decreasing for increasing coupling.
In particular, direct calculation\cite{MPS} of $k_{F} \xi_{{\rm pair}}$ yields
the values $(15.2,3.4.1.1,0.6)$ for
$(k_{F} a_{F})^{-1} = (-2.0,-1.0,0.0,+1.0)$, respectively, in agreement with 
the values extracted from Fig.~7.

We recall at this point that (apart from a numerical factor of order unity)
$\xi_{{\rm phase}}$ and $\xi_{{\rm pair}}$ coincide with
each other in the intermediate-to-weak-coupling side of the crossover
(which is mostly relevant to the BCS contribution).
We thus conclude from the present analysis of the current correlation function
that {\em it is the length} $\xi_{{\rm phase}}$ {\em to determine the range of
the Pippard-like kernel over the whole crossover range}. In addition,
the present discussion shows the importance of including both BCS and AL
type contributions in the intermediate-coupling region, for a correct 
description of the current correlation function.

\section{Concluding Remarks}

In this paper, we have calculated  the current correlation function within the 
BCS-BE crossover. We have complemented the standard analysis of the BCS
diagrammatic contribution to the
current correlation function in the broken-symmetry phase, by considering
an additional (Aslamazov-Larkin
type) diagrammatic contribution which includes fluctuation effects in weak
coupling and recovers
the Bogoliubov form of the current correlation function in strong coupling.

In view of application to high-temperature cuprate superconductors, the
analysis presented in this paper is necessarily a preliminary one, 
since it relies on
a continuum model in three dimensions and does not take into account
additional effects (such as the Coulomb
repulsion between charged composite bosons), which modify the Bogoliubov
picture for bosons.
In addition, lattice effects and the $d$-wave character of the gap function
should also be included in a
realistic calculation of the superfluid density.

The analysis presented in this paper is, nonetheless, a required preliminary 
step toward a more realistic theory of the current correlation function in the
superconducting phase, by connecting results valid in
the weak- and strong-coupling limits in the spirit of the
BCS-BE crossover.
In this context, the identification of the Aslamazov-Larkin type diagram(s)
in the broken-symmetry phase as the dominant one(s) in the strong-coupling
limit of the theory represents one of the main contributions of the present 
paper.
In addition, consideration of the evolution of the current correlation 
function in the broken-symmetry phase from weak to strong coupling has 
required us to deal with
several relevant theoretical issues, such as the mapping between the 
fermionic and bosonic diagrammatic structures in the broken-symmetry phase,
the connection between the Bogoliubov propagators for composite bosons
and the diagrammatic structure for the constituent fermions, and the 
identification of the local and nonlocal parts of the ${\bf j}$ vs ${\bf A}$
relation for a generic diagrammatic contribution to the current correlation 
function.  
The temperature dependence of the superfluid density that we have obtained
(by including the AL type diagram
on top of the conventional BCS ``bubble'' diagram with an $s$-wave gap
function) has served to identify the temperature
regions where one of the two contributions dominates over the other one in
different coupling regimes.
In this context, we have shown that, in the intermediate-coupling region,
superfluidity is most robust against suppressing
mechanisms, such as pair breaking (coming from the weak-coupling side) and
collective excitations (coming from the
strong-coupling side), since the effects of these mechanisms is
considerably reduced in the intermediate-coupling
region.
It would accordingly be tempting to relate this effect to the occurrence
of a maximum for the superconducting critical
temperature in the phase diagram of cuprate superconductors at intermediate
doping. We have also been able to extract from our (zero-temperature) 
calculation of the spatial dependence of the current response kernel  a 
length scale previously introduced in the theory of the BCS-BE crossover, 
which specifies the correlation among different fermion pairs.

Although detailed comparison with experimental data on cuprate
superconductors has to await more realistic calculations
that include lattice effects and the $d$-wave character of the gap
function (for instance, via an extended negative-$U$ Hubbard model), the 
\emph{trend\/} of the coherence length in the Pippard-like kernel to
decrease initially with increasing coupling from
weak to intermediate should be experimentally accessible for cuprate
superconductors upon decreasing the doping (from
overdoping to underdoping), in analogy with the original experimental
results by Pippard \cite{Pippard}.

Previous calculations of the temperature dependence of the superfluid density
have included the $d$-wave character of the gap function 
within the standard BCS ``bubble'' contribution\cite{SD-d-wave-2}, or have 
considered incoherent pair excitations of finite momentum which assist
the destruction of superconductivity (still taking into account the $d$-wave
gap) in the context of the BCS-BE crossover\cite{Levin-rho}. As 
both works do not address the role of collective modes 
(over and above the quasi-particle contribution) in the context of this 
crossover, direct comparison with our work cannot be made.
Additional inclusion of long-range Coulomb interaction addressed in 
Ref.~\onlinecite{plasmon} is more directly relevant to the present approach 
and is presently being considered.

\acknowledgments
The authors are indebted to C. Castellani for discussions.
Financial support from the Italian MIUR under contract COFIN 2001, Prot. 
2001023848 is gratefully acknowledged.

\appendix
\section{Particle-particle ladder propagator in the
broken-symmetry phase and the
Bogoliubov propagators}

In this Appendix, we study in detail the particle-particle ladder 
entering the fermionic diagrammatic structure for the broken-symmetry 
phase, and specifically the AL diagram(s) considered in Section III.
To this end, we shall take the fermionic attractive interaction of the form of
a contact-potential with a suitable regularization, as it was done in 
Ref.~\onlinecite{Pi-S-98} for the normal phase.
In addition, in the strong-coupling limit we establish a connection between the
particle-particle ladder propagator and the Bogoliubov propagators for
bosons, correcting at the same time some erroneous conclusions reached in
 Ref.~\onlinecite{Haussmann} in this context.

Knowledge of the detailed form of the attractive
interaction $V({\bf r})$ of Eq.~(\ref{csibcs}) is not required when studying 
the BCS-BE crossover.
One may accordingly consider the simple form of a ``contact'' potential
$V({\bf r}) = v_{0}  \delta ({\mathbf r})$, where $v_{0}$ is a negative 
constant.
This choice of the potential entails a suitable regularization, e.g., in terms
of a cutoff $k_{0}$ in wave-vector space.
In three dimensions, this can be achieved via the scattering length $a_{F}$
of the associated two-body (fermionic) problem, by choosing the constant 
$v_{0}$ as follows \cite{Pi-S-98}:
\begin{equation}
v_{0}  =  -  \frac{2 \pi^{2}}{m k_{0}}  - 
           \frac{\pi^{3}}{m a_{F} k_{0}^{2}}  \,  .
\label{v0}
\end{equation}
In this way, the classification of the (fermionic) many-body diagrams is
considerably simplified even in the broken-symmetry phase, since only specific
diagrammatic substructures survive when the limit 
$k_{0} \rightarrow \infty$ is eventually taken.

\begin{figure}
\centerline{\epsfig{figure=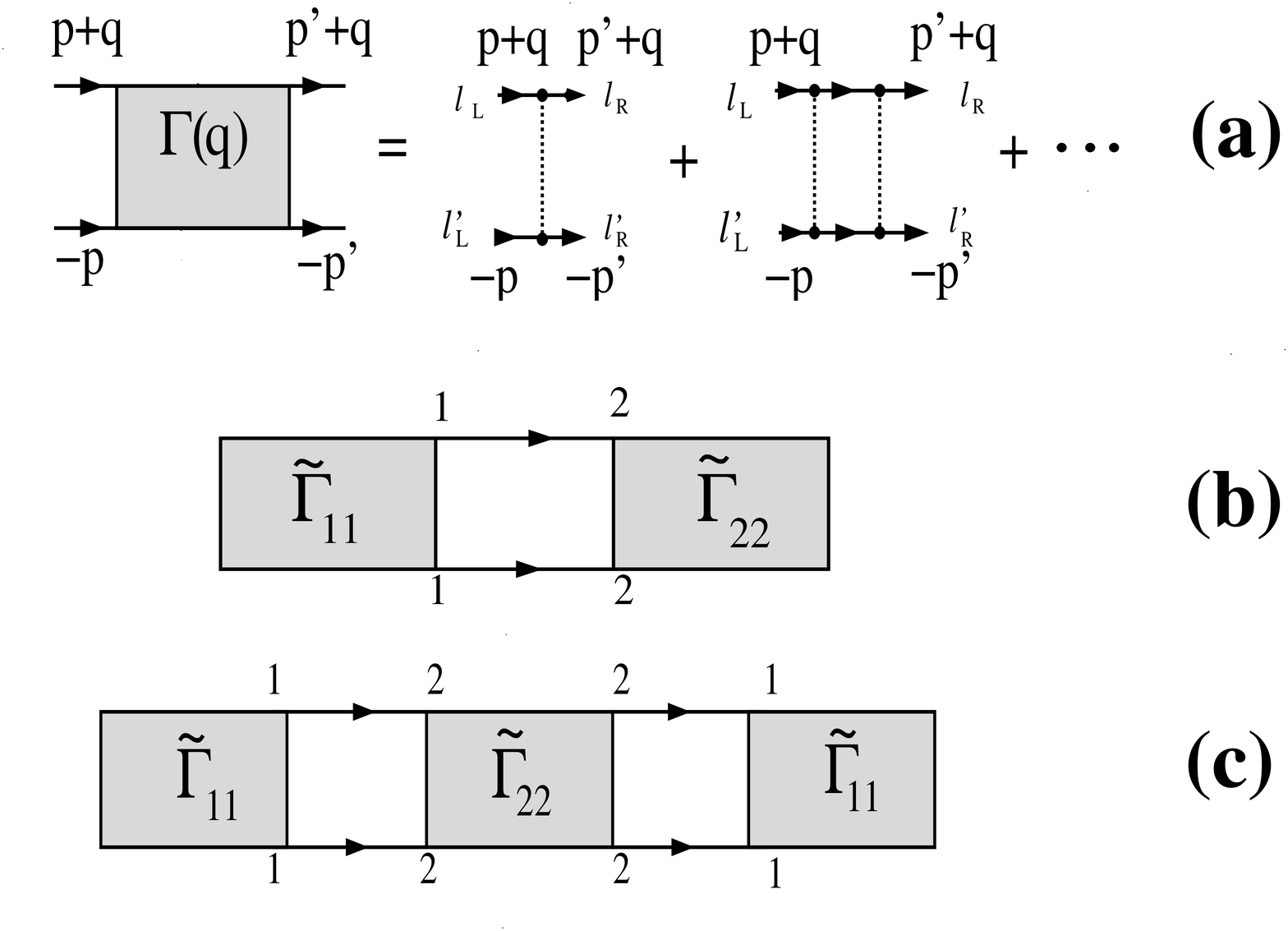,width=6.3cm}}
\caption{(a) Particle-particle ladder in the broken-symmetry
phase (conventions for four-momenta and Nambu indices are specified); (b) 
and (c) show typical insertions of the rungs $\Pi_{12}$ and $\Pi_{21}$, which 
are compatible with the
regularization (\ref{v0}) of the fermionic interaction potential. [Dots 
delimiting the potential (broken line) here represent $\tau_{3}$ Pauli matrices.]}
\end{figure}

With these premises, we consider the \emph{particle-particle ladder\/}
depicted in Fig.~8(a).\cite{footnote-Nambu-arrows}
One can convince oneself that, in these diagrams, the elementary rung
\begin{equation}
K_{\ell_{L} \ell'_{L},\ell_{R} \ell'_{R}}(q)  =  \int \! \frac{d
{\mathbf p}}{(2\pi)^{3}} 
\frac{1}{\beta}  \sum_{n}  {\mathcal G}_{\ell_{R}\ell_{L}}(p+q) 
        {\mathcal G}_{\ell'_{R}\ell'_{L}}(-p)\;,
\label{elementary-rung}
\end{equation}
with the single-particle Green's functions of the BCS form
(\ref{BCS-Green-function}),
survives the regularization introduced by the potential (\ref{v0}) in
the limit
$k_{0} \to \infty$ (and thus $v_0 \to 0$) {\em only\/} if
$\ell_{R}=\ell_{L}=\ell'_{R}=\ell'_{L}=1$ or
$\ell_{R}=\ell_{L}=\ell'_{R}=\ell'_{L}=2$.
[We use the coventions of Fig.~1 for the Nambu indices.]
We then isolate the ultraviolet divergence of (\ref{elementary-rung}) as 
$k_{0} \rightarrow \infty$, by writing
[cf. Eq.~(\ref{Pi-11})]
\begin{equation}
\Pi_{11}(q)  \equiv  K_{11,11}(q)  =  R_{11}(q)  +  \frac{m
k_{0}}{2 \pi^{2}}
\label{defin-Pi-11}
\end{equation}
as well as
\begin{equation}
\Pi_{22}(q)  \equiv  K_{22,22}(q)  =  \Pi_{11}(-q)  \, ,
\label{defin-Pi-22}
\end{equation}
where
\begin{equation}
R_{11}(q)  =  \int \! \frac{d {\mathbf p}}{(2\pi)^{3}}  \left[
\frac{1}{\beta}  \sum_{n}
 {\mathcal G}_{11}(p+q)  {\mathcal G}_{11}(-p)  - 
\frac{m}{{\mathbf k}^{2}} \right] \label{R}
\end{equation}
is a well-behaved quantity in the limit $k_0\to\infty$.
In this way, we are led to introduce the partial ladders
\begin{eqnarray}
\tilde{\Gamma}_{11}(q) & = & -  v_{0}  +  v_{0}^{2}  \Pi_{11}(q)  - 
 v_{0}^{3}  \Pi_{11}(q)^{2}  +  \cdots       \nonumber \\
& = & -  \frac{1}{\frac{m}{4 \pi a_{F}}  +  R_{11}(q)}
\label{partial-Gamma-11}
\end{eqnarray}
and
\begin{eqnarray}
\tilde{\Gamma}_{22}(q) &= &-  v_{0}  +  v_{0}^{2}  \Pi_{22}(q)  - 
v_{0}^{3}  \Pi_{22}(q)^{2}  +  \cdots \nonumber \\
& = & \tilde{\Gamma}_{11}(-q)
 .
\label{partial-Gamma-22}
\end{eqnarray}
This does not imply, however, that in Fig.~8(a) there cannot occur
structures with the pair  $\ell_{L}=\ell'_{L}$ different from the 
pair $\ell_{R}=\ell'_{R}$.
The rungs $\Pi_{12}(q) \equiv K_{11,22}(q)$ or $\Pi_{21}(q) \equiv
K_{22,11}(q)$
[cf. Eq.~(\ref{Pi-12})] can, in fact, be inserted in between the above
partial resummations
$\tilde{\Gamma}(q)_{11}$ and $\tilde{\Gamma}(q)_{22}$, as indicated in
Figs.~8(b) and 8(c).
It is possible to take into account these structures in a systematic way,
by introducing the matrix
equation
\begin{eqnarray}
& & \left( \begin{array}{cc} \Gamma_{11}(q) & \Gamma_{12}(q) \\ \Gamma_{21}(q)
& \Gamma_{22}(q) \end{array} \right)
 =  -  v_{0} \left( \begin{array}{cc} 1 & 0 \\ 0 & 1 \end{array}
\right)   \label{Gamma-matrix-algebraic} \\
& & - v_{0}  \left( \begin{array}{cc} \Pi_{11}(q) & \Pi_{12}(q) \\
\Pi_{21}(q) & \Pi_{22}(q) \end{array} \right)
\left( \begin{array}{cc} \Gamma_{11}(q) & \Gamma_{12}(q) \\
\Gamma_{21}(q) & \Gamma_{22}(q) \end{array} \right) \nonumber  ,
\end{eqnarray}
whose solution yields
\begin{eqnarray}
\left( \begin{array}{cc} \Gamma_{11}(q) & \Gamma_{12}(q) \\ \Gamma_{21}(q)
& \Gamma_{22}(q) \end{array} \right)
& = & \frac{1}{A(q)  A(-q)  -  B(q)^{2}} \nonumber \\
&\times &\left( \begin{array}{cc} A(-q) & B(q) \\ B(q) & A(q) \end{array} \right)
\label{Gamma-solution}
\end{eqnarray}
with the short-hand notation
\begin{eqnarray}
-  A(q) & = & \frac{1}{v_{0}}  + \Pi_{11}(q) \nonumber \\
& = &  \int \! \frac{d
{\mathbf p}}{(2\pi)^{3}}
\left[ \frac{1}{\beta}  \sum_{n}  {\mathcal G}_{11}(p+q) 
{\mathcal G}_{11}(-p)  - 
\frac{1}{2 E({\mathbf p})} \right]
\nonumber \\
B(q) & = & \Pi_{12}(q) .
\label{A-definition}
\end{eqnarray}
Note that in Eq.~(\ref{A-definition}) the gap equation
\begin{equation}
-\frac{1}{v_0}= \int \! \frac{d{\mathbf p}}{(2\pi)^{3}}
\frac{1}{2 E({\mathbf p})}
\end{equation}
has been taken into account.
In this way, we can reproduce all diagrams in Fig.~8 that survive the
regularization of the potential.

The expressions (\ref{A-definition}) for $A(q)$ and $B(q)$ simplify 
considerably {\em in the strong-coupling\/} ($\beta \mu \rightarrow - \infty$)
{\em limit\/} we are interested in. One obtains~\cite{MPS}:
\begin{equation}
A(q)  \simeq  \frac{m a_{F}}{4 \pi}  \left( \frac{a_{F} k_{F}^{3}}{3
\pi}  - 
i  \omega_{\nu}  \frac{m}{2}  +  {\mathbf q}^{2}  \frac{1}{8}
\right)    \label{A-approx}
\end{equation}
and
\begin{equation}
B(q)  \simeq  \frac{m a_{F}^{2} k_{F}^{3}}{12 \pi^{2}}
\label{B-approx}
\end{equation}
to the lowest significant order (where the Fermi wave vector $k_{F}$ is
related to the density $n$
via $n=k_{F}^{3}/(3\pi^{2})$).
This gives for the matrix elements (\ref{Gamma-solution}) the expressions:
\begin{equation}
\Gamma_{11}(q)  =  \Gamma_{22}(-q)  \simeq  \frac{8 \pi}{m^{2}
a_{F}} 
\frac{  \mu_{B}  +  i \omega_{\nu}  +  {\mathbf q}^{2}/(4m)}
{E_{B}({\mathbf q})^{2}  -  (i \omega_{\nu})^{2}}
\label{Gamma-11-approx}
\end{equation}
and
\begin{equation}
\Gamma_{12}(q)  =  \Gamma_{21}(q)  \simeq  \frac{8 \pi}{m^{2} a_{F}} 
\frac{\mu_{B}}{E_{B}({\mathbf q})^{2}  -  (i \omega_{\nu})^{2}}
\label{Gamma-12-approx}
\end{equation}
where
\begin{equation}
E_{B}({\mathbf q})  =  \sqrt{ \left( \frac{{\mathbf q}^{2}}{2 m_{B}}  +
 \mu_{B}\right)^{2} -  \mu_{B}^{2}}
\label{Bogoliubov-disp}
\end{equation}
$m_{B}=2m$ being the bosonic mass and $\mu_{B} = 2 k_{F}^{3} a_{F}/(3 \pi
m)$ the bosonic
chemical potential.
Note that $\mu_{B}$ can be cast in the Bogoliubov form
\begin{equation}
\mu_{B}  =  n_{B}  v_{2}(0)  =  \frac{n}{2}  \frac{4 \pi
a_{F}}{m}    \label{pot-chim-Bog}
\end{equation}
in terms of the bosonic density $n_{B}=n/2$ and of the residual bosonic
interaction $v_{2}(0)=4 \pi a_{F}/m$.\cite{Haussmann,Pi-S-98}
The expression (\ref{Bogoliubov-disp}) has thus the form of the Bogoliubov
dispersion relation \cite{FW}.

The results (\ref{Gamma-11-approx}) and (\ref{Gamma-12-approx}) bear strong
resemblance with
the normal (${\mathcal G}'$) and anomalous (${\mathcal G}_{21}'$)
noncondensate bosonic Green's
functions, respectively, within the Bogoliubov approximation, which read
\cite{FW}:
\begin{eqnarray}
{\mathcal G}'(q) &=&  \frac{\mu_{B} + i \omega_{\nu}  +  
{\mathbf q}^{2}/(2m_{B})}
{(i \omega_{\nu})^{2}  - E_{B}({\mathbf q})^{2}}     \label{Bog-normal}\\
{\mathcal G}_{21}'(q) &=&  -  \frac{\mu_{B}}{(i \omega_{\nu})^{2}  -
 E_{B}({\mathbf q})^{2}} \,  .
\label{Bog-anomalous}
\end{eqnarray}
Comparison of Eqs.(\ref{Gamma-11-approx})-(\ref{Gamma-12-approx}) with
Eqs.~(\ref{Bog-normal})-(\ref{Bog-anomalous}) thus lead us, in fact, to
identify:
\begin{equation}
\left\{ \begin{array}{l} \Gamma_{11}(q)  =  -  (8 \pi / m^{2} a_{F})
 {\mathcal G}'(q) \\
\Gamma_{12}(q)  =  (8 \pi / m^{2} a_{F})  {\mathcal G}_{21}'(q)
\end{array} \right.
\label{Gamma-Bogoliubov}
\end{equation}
where the \emph{sign difference\/} between the two expressions must be 
remarked.

Haussmann\cite{Haussmann} had already noted this sign difference and attempted
to get rid of it by fixing
the phase of the BCS gap function at the particular value $\pi/2$.
This prescription appears rather arbitrary,
since the theory cannot depend on the value of the phase of the gap function.
This sign difference is instead genuine, reflecting the fact that the
particle-particle ladder propagator does \emph{not\/} fully
reduce to the Bogoliubov propagators in the strong-coupling limit.

To demonstrate this point explicitly, we resort directly to the definition
of the Bogoliubov propagators \cite{FW}
\begin{equation}
{\mathcal G}'(x,x')  =  -  \langle T_{\tau}[\Psi_{B}(x)
\Psi^{\dagger}_{B}(x')] \rangle 
+  \alpha({\mathbf r})  \alpha^{*}({\mathbf r'})
\label{defin-G}
\end{equation}
and
\begin{equation}
{\mathcal G}'_{21}(x,x')  =  -  \langle T_{\tau}[\Psi^{\dagger}_{B}(x)
\Psi^{\dagger}_{B}(x')] \rangle 
+  \alpha^{*}({\mathbf r})  \alpha^{*}({\mathbf r'})   ,
\label{defin-G'}
\end{equation}
where $x=({\mathbf r},\tau)$ and $\alpha({\mathbf r}) =
\langle \Psi_{B}({\mathbf r}) \rangle$ is the
\emph{condensate amplitude\/} defined in terms of the bosonic field
operator $\Psi_{B}({\mathbf r})$.
The desired connection between fermionic and bosonic quantities then results by
relating the bosonic operator $\Psi_{B}({\mathbf r})$ to its fermionic
counterpart
$\psi_{\sigma}({\mathbf r})$ (where $\sigma = (\uparrow,\downarrow)$ is
the spin projection),
by setting \cite{PS-96}:
\begin{equation}
\Psi_{B}({\mathbf r})  =  \int \! d{\mathbf \rho}  \phi({\mathbf \rho}) 
\psi_{\downarrow}({\mathbf r}-{\mathbf \rho}/2) 
\psi_{\uparrow}({\mathbf r}+{\mathbf \rho}/2)
\label{definition-boson-fermion}
\end{equation}
where $\phi({\mathbf \rho})$ is a suitable function.
On physical grounds, we choose $\phi({\mathbf \rho})$ to be the (normalized) 
bound-state wave function of the associated (fermionic) two-body problem:
\begin{equation}
\phi({\mathbf \rho})  =  \frac{1}{\sqrt{2 \pi a_{F}}} 
\frac{e^{-\rho/a_{F}}}{\rho}
\label{psi-rho}
\end{equation}
with Fourier transform 
\begin{equation}
\phi({\mathbf p})  =  \sqrt{\frac{8 \pi}{a_{F}}} 
\frac{1}{{\mathbf p}^{2}  +  a_{F}^{-2}}  \, .
\label{psi-k}
\end{equation}
Entering the definition (\ref{definition-boson-fermion}) into
Eqs.~(\ref{defin-G}) and (\ref{defin-G'}),
and recalling the definition (\ref{L}) of the two-particle correlation
function, we rewrite
the Bogoliubov propagators in the compact form:
\begin{equation}
{\mathcal G}'({\mathbf r} \tau,{\mathbf r}' \tau')  =  -  \int \!
d{\mathbf \rho}  \int \!
d{\mathbf \rho}'  \phi({\mathbf \rho})  \phi^{*}({\mathbf \rho}') 
L(1,2;1',2')            \label{G-L}
\end{equation}
with the identification $1=({\mathbf r}+{\mathbf \rho}/2,\tau,\ell=1)$,
$2=({\mathbf r}'-{\mathbf \rho}'/2,\tau',\ell=2)$,
$1'=({\mathbf r}-{\mathbf \rho}/2,\tau^{+},\ell=2)$,
and $2'=({\mathbf r}'+{\mathbf \rho}'/2,\tau'^{+},\ell=1)$; as well as
\begin{equation}
{\mathcal G}'_{21}({\mathbf r} \tau,{\mathbf r}' \tau')  =  -  \int \!
d{\mathbf \rho}  \int \!
d{\mathbf \rho}'  \phi^{*}({\mathbf \rho})  \phi^{*}({\mathbf \rho}') 
L(1,2;1',2')         \label{G'-L}
\end{equation}
with the identification $1=({\mathbf r}-{\mathbf \rho}/2,\tau,\ell=2)$,
$2=({\mathbf r}'-{\mathbf \rho}'/2,\tau',\ell=2)$,
$1'=({\mathbf r}+{\mathbf \rho}/2,\tau^{+},\ell=1)$,
and $2'=({\mathbf r}'+{\mathbf \rho}'/2,\tau'^{+},\ell=1)$.
In Fourier space, we write:
\begin{eqnarray}
{\mathcal G}'(q) & = & -  \int \! \frac{d{\mathbf p}}{(2\pi)^{3}} 
\frac{1}{\beta} \sum_{n} 
\int \! \frac{d{\mathbf p}'}{(2\pi)^{3}}  \frac{1}{\beta} \sum_{n'} 
\phi({\mathbf p}+{\mathbf q}/2) \nonumber \\
& \times & \phi({\mathbf p}'+{\mathbf q}/2) 
L^{11}_{22} ({\mathbf p}\omega_{n},{\mathbf p}'\omega_{n'};q)
\label{G'-FT}
\end{eqnarray}
and
\begin{eqnarray}
{\mathcal G}'_{21}(q) & = & -  \int \! \frac{d{\mathbf p}}{(2\pi)^{3}} 
\frac{1}{\beta} \sum_{n} 
\int \! \frac{d{\mathbf p}'}{(2\pi)^{3}}  \frac{1}{\beta} \sum_{n'} 
\phi({\mathbf p}+{\mathbf q}/2) \nonumber \\
&\times & \phi({\mathbf p}'+{\mathbf q}/2)  
L^{21}_{12} ({\mathbf p}\omega_{n},{\mathbf p}'\omega_{n'};q)
\label{G'-12-FT}
\end{eqnarray}
where the conventions for the four-momenta and Nambu indices are as
in Fig.~1.

We can use at this point the expression (\ref{L-T}) for $L$ in terms of the
many-particle $T$-matrix, and limit ourselves to consider the BCS approximation
(\ref{csibcs}) for the
kernel $\Xi$, as obtained from the off-diagonal terms of the BCS
self-energy.
As already stated in Section II, with this restriction only four elements of
the many-particle
$T$-matrix with $\ell_{L}\neq\ell'_{L}$ and $\ell_{R}\neq\ell'_{R}$ survive.
In addition, for a contact potential the $T$-matrix in Fourier space
satisfies Eq.~(\ref{T-matrix-algebraic}), whose solution yields
\begin{eqnarray}
\left( \begin{array}{cc} T_{11}(q) & T_{12}(q) \\ T_{21}(q) & T_{22}(q)
\end{array} \right)
& = & \frac{1}{A(q)  A(-q)  -  B(q)^{2}} \nonumber \\
& \times &
\left( \begin{array}{cc} - A(-q) & B(q) \\ B(q) & - A(q) \end{array}
\right)       \label{T-solution}
\end{eqnarray}
with the same definitions of Eq.~(\ref{A-definition}).

We are specifically interested in the \emph{strong-coupling limit\/} of the
above expressions, whereby the off-diagonal single-particle Green's functions 
can be neglected in comparison to the diagonal ones since $\Delta\ll |\mu|$ 
in this limit.
In this way, the expressions (\ref{G'-FT}) and (\ref{G'-12-FT}) reduce to:
\begin{equation}
{\mathcal G}'(q)  \simeq  -  {\mathcal F}_{2}(q)  + 
{\mathcal F}_{1}(q)^{2}  T_{11}(q)
\label{G'-FT-sc}
\end{equation}
and
\begin{equation}
{\mathcal G}'_{21}(q)  \simeq  {\mathcal F}_{1}(q)  {\mathcal F}_{1}(-q)
 T_{21}(q)
\label{G'-12-FT-sc}
\end{equation}
where we have introduced the notation ($j=1,2$)
\begin{equation}
{\mathcal F}_{j}(q)  =  \int \! \frac{d{\mathbf p}}{(2\pi)^{3}} 
\phi({\mathbf p}+{\mathbf q}/2)^{j}
 \frac{1}{\beta} \sum_{n}  {\mathcal G}_{11}(p+q) 
{\mathcal G}_{11}(-p)    .  \label{def-F-j}
\end{equation}
With the approximate result
\begin{eqnarray}
\frac{1}{\beta} \sum_{n}  {\mathcal G}_{11}(p+q)  {\mathcal G}_{11}(-p)
& \simeq &
\frac{1}{E({\mathbf k})  +  E({\mathbf k}+{\mathbf q})  -  i
\omega_{\nu}} \nonumber \\ & \simeq &
\frac{1}{2  E({\mathbf k})}
\label{G11-G11-approx}
\end{eqnarray}
that holds specifically in the strong coupling-limit, and using the expression
(\ref{psi-k}) for the two-body wave function, Eq.~(\ref{def-F-j}) can be 
readily evaluated to yield:
\begin{equation}
{\mathcal F}_{1}(q)  \simeq  \sqrt{\frac{m^{2}  a_{F}}{8\pi}}
 \, ,\;\;\;\; 
{\mathcal F}_{2}(q)  \simeq  \frac{m  a_{F}^{2}}{4} \,  .
\label{F-approx}
\end{equation}
Neglecting the first term on the right-hand side of Eq.~(\ref{G'-FT-sc})
(which does not contain the polar structure of the $T$-matrix), we obtain 
eventually:
\begin{eqnarray}
{\mathcal G}'(q) & \simeq & \frac{m^{2}  a_{F}}{8\pi}  T_{11}(q)
\nonumber \\
{\mathcal G}'_{21}(q) & \simeq & \frac{m^{2}  a_{F}}{8\pi}  T_{21}(q)
 \, .   \label{G'-FT-T-approx}
\end{eqnarray}
This result show that (apart from a constant prefactor) it is the 
$T$-matrix and not the particle-particle ladder to reduce to the
Bogoliubov propagators in the strong-coupling limit of the fermionic 
theory.
In particular, comparison of Eq.~(\ref{Gamma-solution}) with
Eq.~(\ref{T-solution}) yields
Eqs.~(\ref{T-Gamma}) of the text, thus accounting for the sign difference
noted in Eqs.~(\ref{Gamma-Bogoliubov}).

Finally, it is interesting to evaluate in the strong-coupling limit also the 
condensate amplitude entering Eqs.~(\ref{defin-G}) and (\ref{defin-G'}), with 
the same approximations used above.
From its definition in terms of the bosonic operator
(\ref{definition-boson-fermion}), we obtain
(for a homogeneous system):
\begin{eqnarray}
\alpha & = & \int \! d{\mathbf \rho}\;  \phi({\mathbf \rho}) 
\langle \psi_{\downarrow}({\mathbf r}-{\mathbf \rho}/2) 
\psi_{\uparrow}({\mathbf r}+{\mathbf \rho}/2) \rangle \nonumber \\
& = & \int \! \frac{d{\mathbf p}}{(2\pi)^{3}}  \phi({\mathbf p}) 
\frac{1}{\beta} \sum_{n}  e^{i \omega_{n} \eta}  {\mathcal G}_{12}(p)
\nonumber \\
& \simeq & \frac{\Delta}{2}  \int \! \frac{d{\mathbf p}}{(2\pi)^{3}} 
\frac{\phi({\mathbf p})}{E({\mathbf p})}
 \simeq  \Delta  \sqrt{\frac{m^{2}  a_{F}}{8\pi}}    \,   .
\label{alpha-Delta}
\end{eqnarray}
Note that a similar relation holds (with $\alpha$ in
Eq.~(\ref{alpha-Delta}) replaced by $\sqrt{n/2}$)  when the diagonal
BCS single-particle Green's function ${\mathcal G}_{11}$ of 
Eq.~(\ref{BCS-Green-function}) is used to evaluate the
particle density $n$ in strong coupling and at low temperature~\cite{MPS}.

\section{Mapping of the fermionic onto the bosonic diagrammatic
structure in the broken-symmetry phase}

In Appendix A we have established a connection 
[cf. Eqs.~(\ref{Gamma-Bogoliubov})] between the Bogoliubov propagators 
(\ref{Bog-normal})-(\ref{Bog-anomalous}) and the particle-particle
ladder (\ref{Gamma-11-approx})-(\ref{Gamma-12-approx}), in the
strong-coupling limit.
Since the Bogoliubov results (\ref{Bog-normal})-(\ref{Bog-anomalous}) for
true bosons can be obtained by considering self-energy corrections to the 
free-boson propagator\cite{Beliaev,Popov,FW}, it should also be possible to 
obtain the expressions
(\ref{Gamma-11-approx}) and (\ref{Gamma-12-approx}) for the
particle-particle ladder 
in the strong-coupling limit, by considering bosonic self-energy corrections
(expressed in
terms of the effective bosonic interaction in the normal phase and of the 
condensate
density) to the particle-particle ladder in the normal phase.
In this way, a \emph{mapping\/} between the fermionic and bosonic
diagrammatic structures
is effectively established in the broken-symmetry phase, in a similar
fashion to what
has already been achieved in the normal phase \cite{Pi-S-98}.

We begin by recalling that the Bogoliubov propagators
(\ref{Bog-normal})-(\ref{Bog-anomalous})
in the broken-symmetry phase can be obtained from the free-boson propagator
${\mathcal G}^{(0)}(q) = (i \omega_{\nu} - {\mathbf q}^{2}/(2 m_{B})^{2} +
\mu_{B})^{-1}$,
by considering the following normal and anomalous bosonic self-energy
corrections:
\begin{eqnarray}
\Sigma_{11}^{B}(q) & = & \Sigma_{22}^{B}(-q)  =  n_{0}  (V_B(0)  + 
V_B({\mathbf q})) \nonumber \\
\Sigma_{12}^{B}(q) & = & \Sigma_{21}^{B}(-q)  =  n_{0}  V_B({\mathbf q})
\label{Sigma-Bogoliubov}
\end{eqnarray}
where $V_B({\mathbf q})$ is the Fourier transform of the bosonic interparticle
potential.
Within the same approximation, the (bosonic) chemical potential is given by
\begin{equation}
\mu_{B}  =  \Sigma_{11}^{B}(0)  -  \Sigma_{12}^{B}(0)  =  n_{0}
 V_B(0)  \, ,
\label{Hugenholtz-Pines}
\end{equation}
in agreement with the Hugenholtz-Pines theorem \cite{FW}.

To obtain the corresponding self-energy corrections for composite
bosons, it is sufficient to
consider the rungs $\Pi_{11}(q)$ [cf. Eq.~(\ref{Pi-11})] and $\Pi_{12}(q)$
[cf. Eq.~(\ref{Pi-12})]
in the strong-coupling limit.
At the lowest significant order, we obtain for the contributions
proportional to $\Delta^{2}$:
\begin{eqnarray}
\Pi_{11}(q) & \rightarrow & -  n_{0}  m  a_{F}^{2}
\nonumber \\
\Pi_{12}(q) & \simeq & \frac{n_{0}}{2}  m  a_{F}^{2}
\label{Pi-approx-sc}
\end{eqnarray}
where $n_{0} = |\alpha|^{2}$ is the condensate density.
Recalling that in the normal phase the particle-particle ladder 
reduces to
\cite{Haussmann,Pi-S-98}
\begin{equation}
\Gamma^{0}(q)  \simeq  -  \left(\frac{8 \pi}{m^{2}  a_{F}}\right) 
\frac{1}{i \omega_{\nu}  -  {\mathbf q}^{2}/(4m)^{2}  +  \mu_{B}}
\label{Gamma-0}
\end{equation}
in the strong-coupling limit, the expression for $\Gamma_{11}$ containing a
single insertion of
$\Pi_{11}$ reads:
\begin{eqnarray}
& \Gamma^{0}(q) & \Pi_{11}(q)  \Gamma^{0}(q)
\nonumber \\
& \simeq & \left( - \frac{8 \pi}{m^{2}a_{F}} \right)  \frac{1}{i
\omega_{\nu} - {\mathbf q}^{2}/(4m)^{2} + \mu_{B}}
 \left( - n_{0}  m  a_{F}^{2} \right) \nonumber \\
&\times &\left( - \frac{8 \pi}{m^{2}a_{F}} \right)  \frac{1}{i \omega_{\nu} -
{\mathbf q}^{2}/(4m)^{2} + \mu_{B}}  \nonumber \\
& = & \left( - \frac{8 \pi}{m^{2} a_{F}} \right)  \frac{1}{i \omega_{\nu}
- {\mathbf q}^{2}/(4m)^{2} + \mu_{B}}
 \left( 2 n_{0} v_{2}(0) \right) \nonumber \\ &\times& \frac{1}{i \omega_{\nu} -
{\mathbf q}^{2}/(4m)^{2} + \mu_{B}}  \label{Gamma-Pi-11}
\end{eqnarray}
where $v_{2}(0) = 4 \pi a_{F}/m$ is the residual bosonic interaction
\cite{Haussmann,Pi-S-98}
introduced in Eq.~(\ref{pot-chim-Bog}).
Apart from the overall factor $- 8 \pi/(m^{2} a_{F})$, the expression
(\ref{Gamma-Pi-11}) for composite
bosons coincides with the lowest-order diagram obtained within the
Bogoliubov approximation for true bosons,
with the self-energy insertion $\Sigma_{11}(q) = 2 n_{0} v_{2}(0)$ (which
coincides with the first
of Eqs.~(\ref{Sigma-Bogoliubov}), provided one approximates $V_B({\mathbf q})
\simeq V_B(0) \rightarrow v_{2}(0)$ for the relevant range of wave vectors).

By a similar token, consideration of the structure of $\Gamma^{0}(q)
\Pi_{12}(q) \Gamma^{0}(-q)$
leads us to identify $\Sigma_{12}(q) = n_{0} v_{2}(0)$ as the off-diagonal
self-energy insertion
for composite bosons to lowest order in $\Delta^{2}$.
[Note that it is just the \emph{sign difference\/} between $\Pi_{11}(q)$ and 
$\Pi_{12}(q)$ in
Eqs.~(\ref{Pi-approx-sc}) which eventually results in the sign difference
between $\Gamma_{11}$
and $\Gamma_{12}$ with respect the corresponding Bogoliubov propagators, as
evidenced in Eqs.~(\ref{Gamma-Bogoliubov}).]

\begin{figure}
\centerline{\epsfig{figure=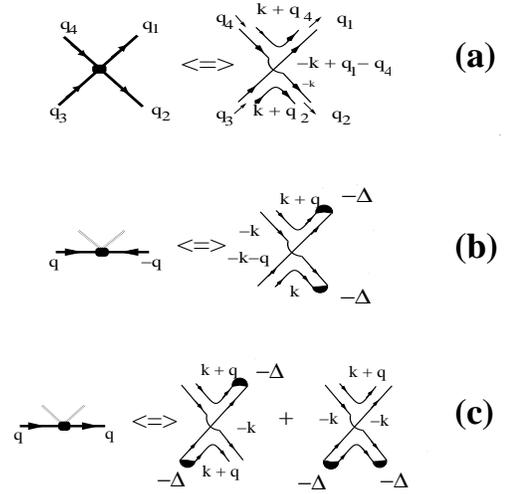,width=6.5cm}}
\caption{Correspondence between the: (a) Effective two-boson
interaction and the four-point fermionic interaction (with 
${\bf q}_1 + {\bf q}_2={\bf q}_3 + {\bf q}_4$); (b) Anomalous and (c) 
normal self-energies within the Bogoliubov approximation and the contractions 
of the four-point fermionic interaction of (a).}
\end{figure}

The above results can be interpreted \emph{euristically\/} in terms of an
effective interaction between composite bosons in the broken-symmetry phase, 
as follows.
Let's consider the four-point vertex for composite bosons \emph{in the
normal phase\/} depicted in Fig.~9(a),
which plays the role of the effective boson-boson 
interaction\cite{Haussmann,Pi-S-98} in the
strong-coupling limit (we shall consider a definite choice of the spin 
indices).

To obtain the bosonic self-energy $\Sigma_{12}^{B}$ of Fig.~9(b) one should:
($i$) contract the two pairs of ``outgoing'' connections in Fig.~9(a), by
joining them through
multiplication with the BCS value of $- \Delta$;
($ii$) multiply each fermionic single-particle line proceeding in the
opposite direction (with respect to one pair of ``ingoing'' connections) by 
$-1$, thus matching the standard convention
[cf. the first of Eqs.~(\ref{BCS-Green-function})].
The final result corresponds to $\Pi_{12}(q)$ of Eq.~(\ref{Pi-approx-sc}), 
since (to the lowest order in $\Delta/|\mu|$) ${\mathcal G}_{12}$ can be 
written in the form:
\begin{equation}
{\mathcal G}_{12}(q)  \simeq  \frac{1}{i \omega_{\nu} - \xi({\mathbf k})}
 ( - \Delta) 
                                 \frac{(-1)}{-i \omega_{\nu} -
\xi({\mathbf k})} \,  .     \label{G-12-G-0}
\end{equation}
Note that in the above equation $\Delta$ plays the role of the square root
of the condensate density
(apart from a numerical coefficient), in agreement with 
Eq.~(\ref{alpha-Delta}).

By a similar token, to obtain the bosonic self-energy $\Sigma_{11}^{B}$ of
Fig.~9(c) one should:
($i$) contract one pair of ``outgoing'' and one pair of
``ingoing'' connections in Fig.~9(a),
by joining them through multiplication with the BCS value of $- \Delta$ in
\emph{two possible ways\/};
($ii$) multiply again each fermionic single-particle line proceeding in the
opposite direction (with respect
to one pair of ``ingoing'' connections) by $-1$.
In the strong-coupling limit, the final result corresponds to 
$-2 \Pi_{11}(q=0)$ of Eq.~(\ref{Pi-approx-sc}).

\section{Extracting the Pippard kernel for bosons within the Bogoliubov 
approximation}

In this Appendix, we obtain the (wave-vector and frequency
dependent)
bosonic current correlation function within the Bogoliubov approximation,
to which
the asymptotic form of the fermionic current correlation function in the
strong-coupling
limit needs be compared.
We shall obtain that by relying on a diagrammatic decoupling method in the
broken-symmetry phase \cite{Bassani-01}.
The static expression of this correlation function is then analyzed to extract
the nonlocal part of the ${\bf j}$ vs ${\bf A}$ relation (Pippard kernel). 
Finally, the penetration depth obtained from Bogoliubov theory is briefly 
discussed.

In analogy to Eq.~(\ref{chi-L}) for fermions, the following expression
holds for the bosonic current correlation function:
\begin{eqnarray}
\chi^{B}_{\gamma, \gamma'}({\mathbf r}\tau,{\mathbf r'}\tau')&=& 
\frac{1}{(2m_{B})^{2}}  (\nabla_{\gamma}  - 
\nabla^{''}_{\gamma}) 
(\nabla^{'}_{\gamma'}  -  \nabla^{'''}_{\gamma'}) 
\label{chi-bosons} \\
&\times&  
\langle T_{\tau} [ \Psi_{B}^{\dagger}({\mathbf r}\tau^{+}) 
\Psi_{B}({\mathbf r''}\tau)\nonumber\\ 
&\times&\Psi_{B}^{\dagger}({\mathbf r'}\tau'^{+})  
\Psi_{B}({\mathbf r'''}\tau') ] \rangle
\vert_{{\mathbf r''}={\mathbf r},{\mathbf r'''}={\mathbf r'}}
\nonumber
\end{eqnarray}
in terms of the bosonic field operator $\Psi_{B}({\mathbf r})$ introduced in
Appendix A.
Setting further
\begin{equation}
\Psi_{B}({\mathbf r})  =  \alpha({\mathbf r})  +  \varphi({\mathbf r})
\label{Psi-splitting}
\end{equation}
where $\alpha({\mathbf r}) = \langle\Psi_B({\mathbf r})\rangle$ is the 
condensate amplitude
and $\varphi({\mathbf r})$ the deviation operator \cite{FW}, the right-hand 
side of Eq.~(\ref{chi-bosons}) is equivalent to the sum
of 16 time-ordered products.
With the factorization
\begin{eqnarray}
& & \langle T_{\tau}[\varphi(1)  \varphi(2)  \varphi^{\dagger}(1') 
\varphi^{\dagger}(2')] \rangle  \nonumber \\
& & = \langle T_{\tau}[\varphi(1)  \varphi(2)] \rangle \langle 
T_{\tau}[\varphi^{\dagger}(1') 
\varphi^{\dagger}(2')] \rangle\nonumber \\
& & +  \langle T_{\tau}[\varphi(1)  \varphi^{\dagger}(1')] \rangle 
\langle T_{\tau}[\varphi(2)  \varphi^{\dagger}(2')] \rangle \nonumber \\
& & + \langle T_{\tau}[\varphi(1)  \varphi^{\dagger}(2')] \rangle 
\langle T_{\tau}[\varphi(2)  \varphi^{\dagger}(1')] \rangle \label{factorization}
\end{eqnarray}
which holds for a system of non-interacting (bosonic) quasi-particles (as
described by the Bogoliubov
approximation), and restricting to the case of a \emph{uniform
condensate\/} whereby $\alpha({\mathbf r}) = \alpha$,
the Fourier transform Eq.~(\ref{chi-bosons}) becomes eventually:
\begin{eqnarray}
& &\chi^{B}_{\gamma, \gamma'}({\mathbf Q},\Omega_{\nu}) =  \int_{0}^{\beta}
 d(\tau-\tau')  e^{i \Omega_{\nu}(\tau-\tau')} 
\int  d({\mathbf r}-{\mathbf r'}) 
\nonumber \\
& & \phantom{\chi^{B}_{\gamma, \gamma'}({\mathbf Q},\Omega_{\nu})} \times  e^{-i {\mathbf Q} \cdot
({\mathbf r}-{\mathbf r'})}\chi^{B}_{\gamma, \gamma'}({\mathbf r}\tau,{\mathbf r'}\tau')
\nonumber \\
& & =  Q_{\gamma}  Q_{\gamma'}  \frac{\alpha^{2}}{(2m_{B})^{2}} 
\left[ {\mathcal G}'(Q)  +  {\mathcal G}'(-Q) 
-  2  {\mathcal G}'_{21}(Q) \right] \nonumber\\ 
& &- \frac{1}{(2m_{B})^{2}} \int  \frac{d{\mathbf q}}{(2\pi)^{3}}  
 (2q_{\gamma}+Q_{\gamma}) (2q_{\gamma'}+Q_{\gamma'})\frac{1}{\beta}  \sum_{\omega_{\nu}}
\nonumber \\
& & \times \left[ {\mathcal G}'(q)  {\mathcal G}'(q+Q) - {\mathcal G}_{21}'(q)  {\mathcal G}_{21}'(q+Q) \right]  .
\label{chi-final}
\end{eqnarray}
Utilizing at this point the form (\ref{Bog-normal} ) and
(\ref{Bog-anomalous}) of the Bogoliubov propagators, the result 
(\ref{chi-final}) is seen to coincide with the
one obtained alternatively by exploiting the Lehmann representation of the
current correlation
function \cite{Bassani-01}.
Although corrections to the Bogoliubov approximation for the current
correlation function
(as to make the longitudinal f-sum rule satisfied while preserving the form
of the Bogoliubov
propagators) have been considered in a formal way \cite{DP-T}, in this
paper we restrict ourselves realistically to the simplest approximation 
(\ref{chi-final}) for the bosonic limit, since we are ultimately interested 
in the evolution of the \emph{fermionic\/} current correlation function
from weak to strong coupling.

The ``static'' $({\mathbf Q} \rightarrow 0,\Omega_{\nu} = 0)$ limit of
Eq.~(\ref{chi-final}) can be used
to test explicitly the longitudinal f-sum rule (\ref{SR}) and to obtain the
(temperature dependence
of the) superfluid density via Eq.(\ref{superfluid-density}).
One obtains:
\begin{eqnarray}
\chi^{B}_{\gamma, \gamma'}({\mathbf Q} \rightarrow 0,\Omega_{\nu} = 0) && \nonumber \\
= -  \frac{n_{0}}{2m_{B}^{2}} && \lim_{{\mathbf Q} \to 0}  Q_{\gamma}  Q_{\gamma'} 
\frac{\left( u_{B}({\mathbf Q})  +  v_{B}({\mathbf Q}) \right)^{2}}{E_{B}({\mathbf Q})}       \nonumber \\
 + \int   \frac{d{\mathbf q}}{(2\pi)^{3}} & &
\frac{q_{\gamma}}{m_{B}}  \frac{q_{\gamma'}}{m_{B}}  \frac{ \partial
n_{B}(E_{B}({\mathbf q})) }
{\partial E_{B}({\mathbf q})}
\label{current-current-final-S-L}
\end{eqnarray}
where
$n_B(E)=(e^{\beta E}-1)^{-1}$ is the Bose-Einstein distribution function, 
$2v_{B}({\mathbf q})^{2}=[({\mathbf q}^{2}/(2m_{B})+
n_{0}V_B({\mathbf q}))/
E_{B}({\mathbf q}) - 1]$ and
$u_{B}({\mathbf q})^{2} - v_{B}({\mathbf q})^{2}  = 1$, $V_B({\mathbf q})$
here being the same bosonic
potential of Appendix B.
Recalling that $v_{B}({\mathbf q})^{2} \simeq m s / (2 |{\mathbf q}|)$  and
$E_{B}({\mathbf q}) \simeq s |{\mathbf q}|$ in the small-${\mathbf q}$ limit
(where $s=\sqrt{n_{0}V_B(0)/m_{B}}$
is the sound velocity), the first term on the right-hand side of
Eq.~(\ref{current-current-final-S-L}) reduces
to $-  \hat{Q}_{\gamma}  \hat{Q}_{\gamma'}  n_{0}/m_{B}$, implying
that the f-sum rule (\ref{SR})
is only approximately verified, insofar as $n_{0} \simeq n_{B}$ within the
Bogoliubov approximation.
The second term on the right-hand side of
Eq.~(\ref{current-current-final-S-L}), on the other
hand, has the form of ($-1/m_{B}$ times) the expression of the Landau's
normal-fluid density $\rho_{n} = n - \rho_{s}$ {(cf., e.g., 
Ref.~\onlinecite{FW}), and is also temperature dependent (albeit through the 
Bogoliubov expression of the quasi-particle excitation energy).

In addition, the Pippard kernel can be obtained from Eq.~(\ref{chi-final})
by letting $\Omega_{\nu}=0$ but keeping ${\mathbf Q}$ finite, in analogy to 
the BCS treatment for fermions\cite{Schrieffer,FW}.
In particular, for a transverse vector potential \emph{in the
zero-temperature limit\/} one considers the expression:
\begin{eqnarray}
& \int \! \frac{d{\mathbf Q}}{(2\pi)^{3}} & \; 
e^{i {\mathbf R}\cdot{\mathbf Q}}\, 
\chi^{B}_{\gamma, \gamma'}({\mathbf Q},\Omega_{\nu}=0)  \nonumber \\
&=& \frac{1}{2 
m_{B}^{2}  (2\pi)^{4}}
\int_{0}^{\infty} \! dq \, q  \int_{0}^{\infty} \! dp  \, p \, G^B(q,p) 
\nonumber \\
& \times &  \lim_{{\mathbf R'} \rightarrow 0}  \frac{\partial}{\partial
R'_{\gamma}} 
\frac{\partial}{\partial R'_{\gamma'}}  j_{0}(q|{\mathbf R'}-{\mathbf R}|)
 j_{0}(p|{\mathbf R'}+{\mathbf R}|)
\label{chi-T=0-finite-Q}
\end{eqnarray}
with the notation $q = |{\bf q}|, p = |{\bf p}|$, and
\begin{eqnarray}
&& G^B(q,p) = \frac{q \, p} {E_{B}(q) 
E_{B}(p)  (E_{B}(q)  +  E_{B}(p))}[ \, \epsilon_{B}(q) \epsilon_{B}(p) \nonumber\\
&& + \, n_{0} V_B(q) \epsilon_{B}(p)  + 
n_{0} V_B(p) \epsilon_{B}(q)-E_{B}(q) E_{B}(p)]
\label{G}
\end{eqnarray}
where $\epsilon_{B}(q)=q^{2}/(2m_{B})$ and $j_{0}(z)=(\sin z)/z$ is the
spherical Bessel function of zeroth order.
The function $G^B(q,p)$ (plotted in Fig.~10) is symmetric under the interchange
of $q$ and $p$, vanishes for $q=p$ but
is elsewhere positive, and its relevant range extends about $q \sim
\xi_{{\rm phase}}^{-1}$ and $p \sim \xi_{{\rm phase}}^{-1}$,
where
\begin{equation}
\xi_{{\rm phase}}  =  \frac{1}{\sqrt{4  m_{B}  n_{0}  V_B(0)}}
\label{xi-phase}
\end{equation}
is the characteristic length of the Bose gas associated with the chemical
potential $\mu_{B} = n_{0} V_B(0)$
[cf. Eq.~(\ref{pot-chim-Bog})] which represents the only energy scale in
the problem \cite{footnote-1}.

\begin{figure}
\centerline{\epsfig{figure=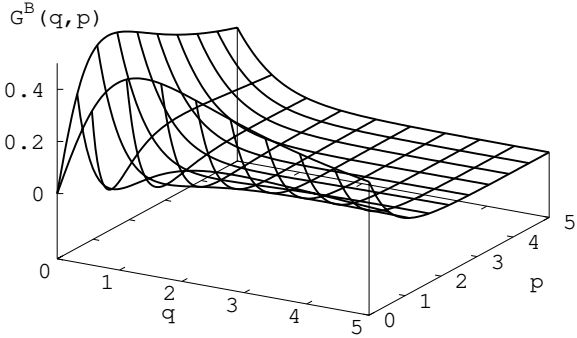,width=6.5cm,angle=-90}}
\caption{Two-dimensional plot of the function $G^B(q,p)$ defined
by Eq.~(\ref{G}), with
the wave vectors in units of $(\sqrt{2}  \xi_{phase})^{-1}$.
\vspace{0.8cm}}
\end{figure}

For the purpose of extracting the leading behavior of the function
(\ref{chi-T=0-finite-Q}) for large values of
$R=|{\mathbf R}|$, the spatial derivatives therein can be calculated with
the same approximations exploited
in the analogous calculation for the BCS (bubble) contribution to the
current correlation function
\cite{Schrieffer,FW}.
One obtains:
\begin{eqnarray}
\lim_{{\mathbf R'} \rightarrow 0}&& \frac{\partial}{\partial R'_{\gamma}} 
\frac{\partial}{\partial R'_{\gamma'}}  j_{0}(q|{\mathbf R'}-{\mathbf R}|)
 j_{0}(p|{\mathbf R'}+{\mathbf R}|)  \nonumber \\
& \approx & -  2  \frac{R_{\gamma}  R_{\gamma'}}{R^{4}}  \cos
(q-p)R  \, .    \label{approx-derivatives}
\end{eqnarray}
Introducing the notation
\begin{equation}
g^B(R)  =  \int_{0}^{\infty} \! dq \, q  \int_{0}^{\infty} \! dp  \, p\,
 G^B(q,p)  \cos (q-p)R \,  ,
\label{g-definition}
\end{equation}
Eq.~(\ref{chi-T=0-finite-Q}) acquires eventually the form:
\begin{eqnarray}
\int \! \frac{d{\mathbf Q}}{(2\pi)^{3}}  \; e^{i {\mathbf R}\cdot{\mathbf Q}}
\, 
\chi^{B}_{\gamma, \gamma'}({\mathbf Q},\Omega_{\nu}=0)  && \nonumber \\
\approx - 
\frac{1}{(2\pi)^{4}  m_{B}^{2}} 
\frac{R_{\gamma}  R_{\gamma'}}{R^{4}}  g^B(R) &&\,  .
\label{chi-final-approx}
\end{eqnarray}
It should be noted that
\begin{equation}
\int_{0}^{\infty} \! dR  \, g^B(R)  =  \pi  \int_{0}^{\infty} \! dq \,
q^{2} \,  G^B(q,q)  =  0
\label{g-integral}
\end{equation}
since $G^B(q,q)=0$ for all values of $q$.

In practice, it is convenient to extract at the ouset a delta-function
contribution from the function
$g^B(R)$, by setting
\begin{equation}
g^B(R)  =  g^B_{n}(R)  -  2  {\mathcal I}^B  \delta(R)
\label{g-decomposition}
\end{equation}
where
\begin{equation}
{\mathcal I}^B  =  \int_{0}^{\infty} \! dR \,  g^B_{n}(R)  \,   ,
\label{I-definition}
\end{equation}
and evaluate the remainder $g^B_{n}(R)$ numerically.
To this end, we exploit the asymptotic form of the function $G^B(q,p)$ and set
\begin{equation}
f^B(\tilde{q})  \equiv  \lim_{\tilde{p} \rightarrow \infty}  \tilde{p}
\,  G^B(\tilde{q},\tilde{p})  = 
2  m_{B}  \frac{ \left( \tilde{q}^{2}  +  1  - 
\sqrt{\tilde{q}^{4}  +  2 \tilde{q}^{2}} \right) }
                   {\sqrt{\tilde{q}^{2}  +  2}}
\label{f-definition}
\end{equation}
where $\tilde{q}=q\sqrt{2}\xi_{{\rm phase}}$ and $\tilde{p}=p\sqrt{2}\xi_{{\rm phase}}$.
With the further notation
\begin{equation}
\bar{G}^B(\tilde{q},\tilde{p})  \equiv  \tilde{q} \,
G^B(\tilde{q},\tilde{p}) \, \tilde{p}  - 
\tilde{q} \, f^B(\tilde{q})  -  \tilde{p} \,  f^B(\tilde{p})   ,
\label{G-bar-definition}
\end{equation}
we obtain:
\begin{eqnarray}\label{g-n-final}
g^B_{n}(R) & = & \frac{1}{4\xi_{{\rm phase}}^{4}} \left\{ \int_{0}^{\infty} \! 
d\tilde{q}  \int_{0}^{\infty} \! d\tilde{p} \,
 \bar{G}^B(\tilde{q},\tilde{p})  \cos (\tilde{q}-\tilde{p})\tilde{R}\right. \nonumber \\
&+& \left.\frac{2}{\tilde{R}} 
\int_{0}^{\infty} \!  d\tilde{q}  \tilde{q}  f^B(\tilde{q})  \sin
(\tilde{q}\tilde{R}) \right\} 
\end{eqnarray}
where $\tilde{R}=R/(\sqrt{2}\xi_{{\rm phase}})$, as well as
\begin{equation}
{\mathcal I}^B  =  -   \frac{\pi\sqrt{2}}{4  \xi_{{\rm phase}}^{3}} 
\int_{0}^{\infty} \! d\tilde{q} 
\,\tilde{q} \, f^B(\tilde{q})  =  -  \frac{\pi  m_{B}}{3 
\xi_{{\rm phase}}^{3}}   \, .               \label{I-value}
\end{equation}
The function $g^B_{n}(R)$ (plotted in Fig.~11) is negative, diverges 
like $m_B(\pi - 1)/(2\xi_{{\rm phase}})\ln \tilde{R}$ for $\tilde{R} \rightarrow 0$, is strongly localized over the range
$\xi_{{\rm phase}}$, and shows minor oscillations about zero for $R>\xi_{{\rm phase}}$.

In contrast to an analogous procedure within the fermionic BCS
approximation of Appendix D,
the strength ${\mathcal I}^B$ of the delta function in
Eq.~(\ref{g-decomposition}) depends on the condensate density via
$\xi_{{\rm phase}}$ (in particular, the strength of the delta function and
consequently the area enclosed by $g^B_{n}$
vanish in the limit $\xi_{{\rm phase}} \rightarrow \infty$, corresponding to
noninteracting (condensed) bosons).
Accordingly, the delta-function contribution \emph{does not cancel\/} the
diamagnetic term in the relation between the
induced current ${\mathbf j}$ and the vector potential ${\mathbf A}$, as it
happens instead for the BCS case \cite{FW} (cf.~also Appendix D).
In the present case, in fact, one obtains:
\begin{eqnarray}
j_{\gamma}({\mathbf r}) & = & -  \frac{1}{m_{B}  c}  \left( n_{B}  - 
\frac{1}{36  \pi^{2}  \xi_{{\rm phase}}^{3}} \right) 
A_{\gamma}({\mathbf r})  \label{j-A-London- Pippard}  \\
& + & \frac{1}{(2\pi)^{4}  m_{B}^{2}  c}  \int \! d{\mathbf r'} 
\frac{ R_{\gamma}  \sum_{\gamma'} R_{\gamma'} A_{\gamma'}({\mathbf r'})
}{R^{4}}  g^B_{n}(R)    \nonumber 
\end{eqnarray}
where ${\mathbf R}={\mathbf r}-{\mathbf r'}$ and $c$ is the light velocity.
Note how the form (\ref{j-A-London- Pippard}) for the current response
kernel is \emph{intermediate\/} between the
(local) London's form and the (nonlocal) Pippard's form \cite{FW}.
Note also that, in the noninteracting-boson limit (whereby $\xi_{{\rm phase}}
\rightarrow \infty$), Eq.(\ref{j-A-London- Pippard})
correctly reduces to the local London's form
\begin{equation}
j_{\gamma}({\mathbf r})  =  -  \frac{n_{B}}{m_{B}  c} 
A_{\gamma}({\mathbf r})   ,        \label{j-A-London}
\end{equation}
in agreement with the result obtained in Ref.~\onlinecite{Schafroth} for a
condensed Bose gas at zero temperature
(for fermions within the BCS approximation, on the other hand, only the
nonlocal Pippard's term survives \cite{FW}).

\begin{figure}
\centerline{\epsfig{figure=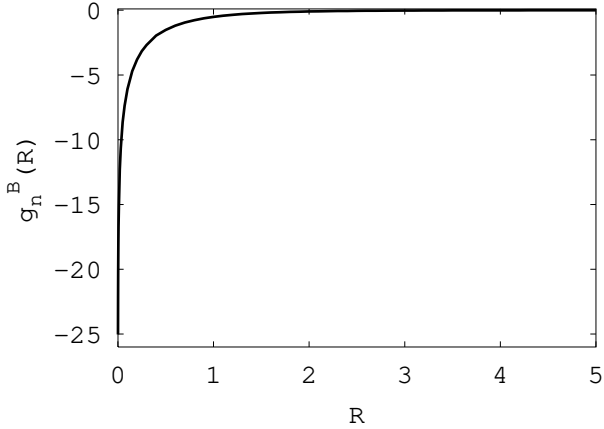,width=6.0cm,angle=-90}}
\caption{Plot of the function $g^B_{n}(R)$ 
(in units of $m_B/(4 \xi_{\rm{phase}}^4)$) as determined
numerically from Eq.~(\ref{g-n-final}),
vs $R$ (in units of $\sqrt{2} \xi_{phase}$).}
\end{figure}

Finally, the real-space relation (\ref{j-A-London- Pippard}) can be 
transposed to a \emph{local condition\/} in wave-vector space as follows.
Taking ${\mathbf q} // \hat{z}$ and ${\mathbf A}({\mathbf q}) // \hat{y}$
(such that
${\mathbf q} \cdot {\mathbf A}({\mathbf q}) = 0$ in the London gauge), we obtain:
\begin{equation}
j_{y}({\mathbf q})  =  -  \frac{c}{4 \pi}  K(q)  A_{y}({\mathbf q})
\label{j-A-q-space}
\end{equation}
where
\begin{equation}
-  \frac{c}{4 \pi}  K(q)  =  -  \frac{1}{m_{B}  c}  \left(
n_{B}  - 
\frac{1}{36  \pi^{2}  \xi_{{\rm phase}}^{3}} \right)  +  Q_{yy}(q)
\label{K}
\end{equation}
and
\begin{equation}
Q_{yy}(q)  =  -  \frac{1}{4  \pi^{3}  m_{B}^{2}  c} 
\int_{0}^{\infty} \! dr 
\frac{g^B_{n}(r)}{r^{2}}  \frac{1}{q}   \frac{\partial}{\partial q} 
\left( \frac{ \sin qr}{qr} \right)   .
\label{Q}
\end{equation}
Following Pippard's treatment \cite{FW,Pippard}, we can then define the
\emph{penetration depth\/} via the
relation:
\begin{equation}
\lambda  =  \frac{2}{\pi}  \int_{0}^{\infty} \! dq 
\frac{1}{q^{2}  +  K(q)}  .      \label{lambda}
\end{equation}
This expression has to be compared with London's relation
at zero temperature
\begin{equation}
\lambda_{L}  =  \sqrt{ \frac{ m_{B}  c^{2} }{ 4\pi  n_{B} 
e_{B}^{2} } }
             =  \sqrt{ \frac{ m  c^{2} }{ 4\pi  n  e^{2} } }
\label{lambda-London}
\end{equation}
(where the electric charge $e_{B}=2 e$ has been restored for convenience).
To this end, we follow an analogous treatment within BCS theory \cite{FW}
and consider the \emph{local \/}
($\lambda \gg \xi_{{\rm phase}}$) and \emph{nonlocal \/} ($\lambda \ll 
\xi_{{\rm phase}}$) limits separately, by
analyzing the $q$-dependence of the kernel (\ref{K}).
Note that in the present case, while the ratio $Q_{yy}(q)/Q_{yy}(0)$ is a
universal function of
$\tilde{q}=q\sqrt{2}\xi_{{\rm phase}}$, the ratio $K(q)/K(0)$ is not, since its
asymptotic limit
$K(\infty)/K(0)=1-(36\pi^{2}n_{B}\xi_{{\rm phase}}^{3})^{-1}$ (with $K(0)=
\lambda_{L}^{-2}$) is nonvanishing.
This contrasts the situation in BCS theory where $K(\infty)=0$,
the difference originating
from the fact that in Bogoliubov theory cancellation of the diamagnetic
term in Eq.~(\ref{K}) does not
occur.
In the local ($\lambda \gg \xi_{{\rm phase}}$) limit, whereby we may 
approximate
$K(q)$ by $K(0)$ in Eq.~(\ref{lambda}),
we obtain $\lambda \simeq \lambda_{L}$.
In the nonlocal ($\lambda \ll  \xi_{{\rm phase}}$) limit, on the other hand,
whereby we may approximate $K(q)$ by
$K(\infty)$ in Eq.~(\ref{lambda}), we obtain $\lambda \simeq \lambda_{L}
(K(\infty)/K(0))^{-1/2} \simeq \lambda_{L}$
owing to the diluteness condition \cite{footnote-2}.
Within the Bogoliubov theory, there is thus no way to get from
Eq.~(\ref{lambda}) values for $\lambda$ significantly
different from London's value $\lambda_{L}$, in contrast with BCS 
theory\cite{FW} whereby one obtains $\lambda\gg\lambda_{L}$ in the nonlocal 
limit.

\section{Extracting the Pippard kernel for superconducting fermions within
the BCS approxmation}

In this Appendix, we set up a general procedure to extract the Pippard
kernel from the BCS bubble contribution to the current correlation function.
Contrary to the method introduced by BCS theory to deal with the (extreme)
weak-coupling limit \cite{FW}, our procedure is applicable to all
couplings.
Although some expressions reported in this Appendix are similar to those of 
Appendix C for the bosonic counterpart, we shall report them here 
to point out similarities and differences between the two cases.

In the BCS method\cite{FW}, a local (delta-function) contribution is 
identified from the paramagnetic contribution to the current induced by a 
(transverse) vector potential.
This local contribution is shown to cancel completely the diamagnetic term, 
thus making ``nonlocal'' the relation between the current and the vector 
potential in coordinate space (over the scale of the Pippard coherence length 
$\xi_{0}$).
We will show in the following that this cancellation of the diamagnetic
term by the local term originating from the paramagnetic contribution is, 
strictly speaking, never complete.
Yet, it can be considered \emph{approximately\/} complete
insofar as the length scale $k_{F}^{-1}$
is much smaller than $\xi_{0}$ (a situation which certainly applies to the
extreme weak-coupling limit).
At stronger coupling (when $k_{F}^{-1}$ becomes comparable with $\xi_{0}$),
on the other hand, this argument is bound to fail.
As a consequence, the relation between the current and the vector potential
in coordinate space contains a nonlocal
as well as a local contribution.

Quite generally, the current correlation function at zero frequency 
obtained from the BCS bubble can be cast in the form:
\begin{eqnarray}
\chi^{BCS}_{\gamma, \gamma'}({\bf Q},\Omega_{\nu}=0) & = & - 
\frac{1}{(2m)^{2}}  \int \! \frac{d {\bf q}}{(2\pi)^{3}} 
(2q_{\gamma} + Q_{\gamma}) \nonumber \\
&\times & (2q_{\gamma'} + Q_{\gamma'}) \frac{G^{BCS}({\bf q},{\bf q}+{\bf Q})}{2  |{\bf q}| 
|{\bf q}+{\bf Q}|}        \label{chi-G}
\end{eqnarray}
where (cf., e.g., Ref.~\onlinecite{FW})
\begin{equation}
G^{BCS}(q,p)  =  -  2  q  p  \, \frac{\xi(q)  \xi(p)  + 
\Delta^{2}  -  E(q)  E(p)}
                                        {E(q)  E(p)  (E(q)  + 
E(p))}  .              \label{G-BCS}
\end{equation}
Note that, similarly to the function $G^{B}(q,p)$ of Eq.~(\ref{G}),  the 
function $G^{BCS}(q,p)$ is also symmetric under the interchange of $p$
and $q$, vanishes for $p=q$,
and is elsewhere positive.

In coordinate space we write correspondingly 
[cf.~Eq.~(\ref{chi-T=0-finite-Q})]:
\begin{eqnarray}
& \int \! \frac{d{\bf Q}}{(2\pi)^{3}} & \; e^{i {\bf R}\cdot{\bf Q}}\, 
\chi^{BCS}_{\gamma, \gamma'}({\bf Q},\Omega_{\nu}=0) \nonumber 
\\ & = & \frac{1}{2  m^{2}  (2\pi)^{4}}
\int_{0}^{\infty} \!dq  \, q  \int_{0}^{\infty} \!dp \, p \,  G^{BCS}(q,p)
\nonumber \\
& \times &  \lim_{{\bf R'} \rightarrow 0}  \frac{\partial}{\partial
R'_{\gamma}} 
\frac{\partial}{\partial R'_{\gamma'}}  j_{0}(q|{\bf R'}-{\bf R}|)
 j_{0}(p|{\bf R'}+{\bf R}|) .
\label{chi-T=0-finite-Qapp}
\end{eqnarray}
As in Appendix C, for the purpose of extracting the leading behavior of the 
function (\ref{chi-T=0-finite-Qapp}) for large values of $R=|{\bf R}|$, we can
make the approximation (\ref{approx-derivatives}) in 
Eq.~(\ref{chi-T=0-finite-Qapp}).

Introducing further the notation [cf.~Eq.~(\ref{g-definition})]:
\begin{equation}
g^{BCS}(R)  =  \int_{0}^{\infty}\!dq\,  q  \int_{0}^{\infty}\!dp\,p \,
 G^{BCS}(q,p)  \cos (q-p)R  ,
\label{g-definitionapp}
\end{equation}
Eq.~(\ref{chi-T=0-finite-Qapp}) acquires eventually the form 
[cf.~Eq.~(\ref{chi-final-approx})]:
\begin{eqnarray}
\int && \frac{d{\bf Q}}{(2\pi)^{3}}  \; e^{i {\bf R}\cdot{\bf Q}} \, 
\chi^{BCS}_{\gamma, \gamma'}({\bf Q},\Omega_{\nu}=0) \nonumber \\
&& \approx  - 
\frac{1}{(2\pi)^{4}  m^{2}} 
\frac{R_{\gamma}  R_{\gamma'}}{R^{4}}  g^{BCS}(R) \, .
\label{chi-final-approxapp}
\end{eqnarray}
It should be noted that also in this case
\begin{equation}
\int_{0}^{\infty} \!dR\, g^{BCS}(R)  =  \pi  \int_{0}^{\infty} \!dq\,q^{2} 
 G^{BCS}(q,q)  =  0
\label{g-integralapp}
\end{equation}
\noindent
since $G^{BCS}(q,q)=0$ for all values of $q$.

The integral in Eq.~(\ref{g-definitionapp}) must be treated with care,
since the product $p\,  G^{BCS}(q,p)$
does not vanish in the limit of large $p$, thus implying a delta-function
contribution to the function $g^{BCS}(R)$ as for the boson case of Appendix C.
To extract this contribution, we exploit the asymptotic form
of the function $G^{BCS}(q,p)$
and set [cf.~Eq.~(\ref{f-definition})]
\begin{equation}
f^{BCS}(q)  \equiv  \lim_{p \rightarrow \infty}  p \,  G^{BCS}(q,p)  =  8  m
 q  v_{q}^{2}   \label{f-definitionapp}
\end{equation}
where $v_{q}^{2}=(1-\xi(q)/E(q))/2$ is the usual BCS factor.
With the further notation
\begin{equation}
\bar{G}^{BCS}(q,p)  \equiv  q \,  G^{BCS}(q,p) \, p  -  q  f^{BCS}(q)  -  p 
f^{BCS}(p)   ,       \label{G-bar-definitionapp}
\end{equation}
Eq.~(\ref{g-definitionapp}) becomes  [cf.~Eq.~(\ref{g-decomposition})]:
\begin{equation}
g^{BCS}(R)  =  g^{BCS}_{n}(R)  -  2  {\mathcal I}^{BCS}  \delta(R)
\label{g-decompositionapp}
\end{equation}
where [cf.~Eq.~(\ref{g-n-final})]
\begin{eqnarray}
g^{BCS}_{n}(R) & = & \int_{0}^{\infty} \!dq  \int_{0}^{\infty} \!dp\,  
\bar{G}^{BCS}(q,p)  \cos (q-p)R \nonumber \\
           & +&  \frac{2}{R}  \int_{0}^{\infty} \!dq\,  q  f(q) 
\sin (q R)           \label{g-n}
\end{eqnarray}
and
\begin{equation}
{\mathcal I}^{BCS}  =  -   \pi  \int_{0}^{\infty} \!dq\,   q  
f^{BCS}(q)  =  -  8  \pi^{3}  m  n  ,
\label{I-valueapp}
\end{equation}
$n$ being the particle density.
Note that ${\mathcal I}^{BCS}$ is a constant, while its bosonic counterpart of 
Eq.~(\ref{I-value}) depend on the bosonic interaction.
In terms of the above quantities, the relation between the induced current
${\bf j}$ and the vector
potential ${\bf A}$ becomes eventually:
\begin{eqnarray}
j_{\gamma}({\mathbf r}) & = & -  \frac{1}{m  c}  \left( n  + 
\frac{{\mathcal I}^{BCS}}{12  \pi^{3}  m} \right)  A_{\gamma}({\bf r})
\label{j-A-general-BCS-ALapp}\\
& + & \frac{1}{(2\pi)^{4}  m^{2}  c}  \int \! d{\bf r'} 
\frac{ R_{\gamma}  \sum_{\gamma'} R_{\gamma'} A_{\gamma'}({\bf r'}) }{R^{4}}
  g^{BCS}_{n}(R) \nonumber
\end{eqnarray}
in the place of Eq.~(\ref{j-A-London- Pippard}) for bosons.
Insertion of the value (\ref{I-valueapp}) into the first term on the
right-hand side of
Eq.~(\ref{j-A-general-BCS-ALapp}) does \emph{not\/} lead to a complete
cancellation of the diamagnetic
term, since $n+{\mathcal I}^{BCS}/(12\pi^{3}m) = n/3$ in this case.
In the standard BCS treatment\cite{FW}, on the other hand, it is claimed that 
the diamagnetic term is completely cancelled by the local (delta-function) 
contribution originating from the paramagnetic term.
This apparent contradiction is resolved by the following argument.

A crucial point in our analysis is that $g_{n}^{BCS}(R)$ is a \emph{regular\/}
function of $R$, in the sense
that it does not contribute any additional delta-function term to the
function $g^{BCS}(R)$, besides the one
explicitly indicated in Eq.~(\ref{g-decompositionapp}).
By a careful analysis of the one- and two-dimensional integrals entering
the definition (\ref{g-n}),
it can, in fact, be shown that $g_{n}^{BCS}(R)$ contains at most the following
(integrable) logarithmic
singularity
\begin{equation}
g_{n}^{BCS}(R)  \approx  -  16  (\pi - 1)  m^{3}  \Delta^{2} \ln
\left( \frac{1}{R  k_{F}} \right)
\label{g-n-singularity}
\end{equation}
for $R  k_{F} \ll 1$.
Note that the form of this singularity is \emph{universal\/}, in the sense
that it depends on the coupling value
only through the value of the gap function $\Delta$ in the prefactor.

To the singular term (\ref{g-n-singularity}) one should add nonsingular
contributions that depend on the value of the coupling parameter.
In particular, for $R  k_{F} \ll 1$ in weak coupling one obtains the
nonsingular contribution $4 m k_{F}^{4}$.
This implies that, in weak coupling, the appearance of the singular term
(\ref{g-n-singularity}) is restricted to
extremely small values of $R$, and it cannot be detected for all
practical purposes.
For instance, when $k_{F} \xi_{0} = k_{F}^{2}/(\pi m \Delta) = 32$ (so that
the extreme weak-coupling limit is definitively reached),
one obtains that the singular term (\ref{g-n-singularity}) dominates over
the constant contribution $4 m k_{F}^{4}$
only for $k_{F} R \lesssim \exp (-1000)$.
At larger couplings, however, the singular term (\ref{g-n-singularity}) 
becomes non negligible for appreciable values of $k_{F} R$, as seen in Fig.~7 
of the text. [We prefer in this Appendix to use the Pippard coherence length 
$\xi_0$ instead of $\xi_{{\rm pair}}$ to make easier the comparison with the
BCS method to obtain the Pippard kernel in the
extreme weak-coupling limit.] 

\begin{figure}
\centerline{\epsfig{figure=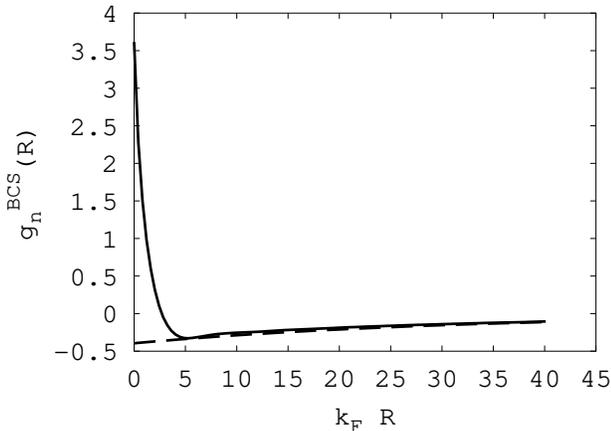,width=6.0cm,angle=-90}}
\caption{Plot of the function $g_{n}^{BCS}(R)$ (in units of $m k_F^4$) as 
determined numerically from Eq.~(\ref{g-definitionapp}),
vs  $k_F R$  for $k_F \xi_0=32$ (full line). The corresponding 
function $-(2 m \pi k_F)^2\Delta\exp(-R/\xi_0)$ is also shown for 
comparison (dashed line).}
\end{figure}

A plot of the function $g_{n}^{BCS}(R)$ (in units of $m k_{F}^{4}$) vs 
$k_{F} R$ obtained by numerical evaluation of
Eq.~(\ref{g-n}) is shown in Fig.~12 (full line) for $k_{F} \xi_{0} = 32$,
whereby the singular behavior
(\ref{g-n-singularity}) is not visible by the above argument.
From this plot, the two characteristic length scales of BCS theory (namely,
$k_{F}^{-1}$ and $\xi_{0}$
such that $k_{F}^{-1} \ll  \xi_{0}$ in the weak-coupling limit) are evident.
Specifically, from the plot of $g_{n}^{BCS}(R)$ one can isolate its ``tail''
represented by the function
$-(2 m \pi k_F)^2\Delta \exp (-R/ \xi_{0})$ 
(or, better, by the function $J(R)$ of BCS theory obtained by integrating the
Bessel function of imaginary argument \cite{FW}), shown in Fig.~12 by the
dashed line.
Once this tail has been isolated, the remainder $g_{n}^{rem}(R) = g_{n}(R)
+(2 m \pi k_F)^2\Delta \exp (-R/ \xi_{0})$ is localized
on the scale $k_{F}^{-1}$, which is much smaller than the scale $\xi_{0}$
of the tail in the weak-coupling
limit.
For these reasons, the \emph{remainder\/} itself can be \emph{effectively
assimilated to a delta-function contribution\/}
with finite strength (by setting $g_{n}^{rem}(R) = - 2 {\mathcal I}'
\delta(R)$), which thus adds up to the true
delta-function contribution of Eq.~(\ref{g-decompositionapp}).
Note that, to cancel completely the diamagnetic term in
Eq.~(\ref{j-A-general-BCS-ALapp}), the factor ${\mathcal I}'$
should equal $- 4 \pi^{3} m n$.
This value is indeed obtained by the data of Fig.~12 within a $10^{-3}$ relative error.

Our procedure has to be contrasted with the standard BCS treatment in the
weak-coupling limit\cite{FW}, whereby
the delta-function contribution that cancels completely the diamagnetic
term in the ${\bf j}$ vs ${\bf A}$
relation is obtained by letting the lower limit of an integration over
$\xi(p)$ and $\xi(q)$ going to
$- \infty$ (instead of holding it at the original finite value $-
k_{F}^{2}/(2m)$).

Although the two procedures yield correctly the same results in the
weak-coupling limit, our procedure
can be readily extended to the intermediate-coupling region where the two
length scales $k_{F}^{-1}$ and
$\xi_{0}$ becomes comparable, as shown in Section IV.

\newpage










\noindent


\begin{references}

\bibitem{Schrieffer} J.R. Schrieffer, \emph{Theory of Superconductivity\/}
                     (W.A. Benjamin, New York, 1964), Chapter 8.

\bibitem{FW} A.L. Fetter and J.D. Walecka, \emph{Quantum Theory of
Many-Particle Systems\/}
            (McGraw-Hill, New York, 1971).

\bibitem{Randeria-90} M. Randeria, J.-M. Duan, L.-Y. Shieh, Phys. Rev. B
{\bf 41}, 327 (1990).

\bibitem{Haussmann} R. Haussmann, Z. Phys. B {\bf 91}, 291 (1993).

\bibitem{PS-94} F. Pistolesi and G.C. Strinati, Phys. Rev. B {\bf 49}, 6356
(1994).

\bibitem{PS-96} F. Pistolesi and G.C. Strinati, Phys. Rev. B {\bf 53},
15168 (1996).

\bibitem{Levin} B. Jank\'o, J. Maly, and K. Levin, Phys. Rev. B {\bf 56},
R11407 (1997).

\bibitem{Zwerger} S. Stintzing and W. Zwerger, Phys. Rev. B {\bf 56}, 9004
(1997).

\bibitem{Pi-S-98} P. Pieri and G.C. Strinati, Phys. Rev. B {\bf 61}, 15370
(2000), and
                  cond-mat/9811166.

\bibitem{Ding} H. Ding \emph{et al.\/}, Nature {\bf 382}, 51 (1996); and
               Phys. Rev. Lett. {\bf 78}, 2628 (1997).

\bibitem{Loeser} A. G. Loeser \emph{et al.\/}, Science {\bf 273}, 325 (1996).

\bibitem{Uemura} Y. J. Uemura \emph{et al.\/}, Phys. Rev. Lett. {\bf 62},
2317 (1989).

\bibitem{Eagles} D.M. Eagles, Phys. Rev. {\bf 186}, 456 (1969).

\bibitem{Leggett} A.J. Leggett, in \emph{Modern Trends in the Theory of
Condensed Matter\/},
                  edited by  A. Pekalski and R. Przystawa, Lecture Notes in
Physics Vol.115
                 (Springer-Verlag, Berlin, 1980), p.13.

\bibitem{NSR} P. Nozi\`{e}res and S. Schmitt-Rink, J. Low. Temp. Phys. {\bf
59}, 195 (1985).

\bibitem{Haussmann-2} R. Haussmann, Phys. Rev. B {\bf 49}, 12975 (1994).

\bibitem{PPSC} A. Perali, P. Pieri, G.C. Strinati, and C. Castellani, 
Phys. Rev. B {\bf 66}, 024510 (2002).

\bibitem{leo} P. Pieri, L. Pisani, and G.C. Strinati (unpublished).

\bibitem{dennis} P. Pieri, G.C. Strinati, and D. Moroni, Phys. Rev. Lett. 
{\bf 89}, 127003 (2002).

\bibitem{SPL} G.C. Strinati, P. Pieri, and C. Lucheroni, Eur. Phys. J. B 
{\bf 30}, 161 (2002).

\bibitem{Luban} M. Luban and W.D. Grobman, Phys. Rev. Lett. {\bf 17}, 182
(1966); see also
                M. Luban in \emph{Quantum Fluids\/}, N. Wiser and D.J.
Amit, Eds.
                (Gordon and Breach, New York, 1970), p.117.

\bibitem{Bassani-GCS} G.C. Strinati, in {\it Electrons and Photons in
Solids, a Volume in honour
                      of Franco Bassani\/} (Scuola Normale Superiore, Pisa,
2001), p.403.

\bibitem{HM} P.C. Hohenberg and P.C Martin, Ann. Phys. {\bf 34}, 291 (1965).

\bibitem{heiselberg} A. Brunello,  F. Dalfovo, L. Pitaevskii, and S. Stringari,
Phys. Rev. Lett. {\bf 85}, 4422 (2000); 
J. M. Vogels, K. Xu, C. Raman, J. R. Abo-Shaeer, and W. Ketterle,
 Phys. Rev. Lett. {\bf 88}, 060402 (2002).

\bibitem{AL} L.G. Aslamazov and A.I. Larkin, Sov. Phys. JETP {\bf 10}, 875
(1968).


\bibitem{Pippard} A.B. Pippard, Proc. Roy. Soc. (London) {\bf A216}, 547 
(1953); T.E. Faber and A.B. Pippard, Proc. Roy. Soc. (London) {\bf A231}, 
336 (1955).

\bibitem{btcc} L. Benfatto, A. Toschi, S. Caprara, and C. Castellani,
Phys. Rev. B {\bf 66}, 054515 (2002).

\bibitem{Baym} G. Baym, Phy. Rev. {\bf 127}, 1391 (1962).

\bibitem{Strinati-RNC} G. Strinati, La Rivista del Nuovo Cimento Vol.{\bf
11}, N.12, (1988).

\bibitem{footnote-Nambu-arrows} The arrows attached to the Nambu Green's 
functions have the meaning of pointing from the second
to the first argument of the Green's functions.
 The distinction between particle-particle and particle-hole diagrams
is, however, purely conventional since particle and hole modes get
intimately interrelated in the broken-symmetry phase.
We shall nevertheless mantain the terminology used for the normal phase
and refer, for instance, to the diagrams of Fig.~8(a) as to the 
``particle-particle'' ladder diagrams.
\bibitem{baymrhos} G. Baym, in {\em Mathematical Methods in Solid State and
Superfluid Theory}, Scottish Universities' Summer School 1967, R.C. Clark and 
G.H. Derrick, Eds. (Oliver \& Boyd, Edinburgh, 1967), p. 121.
\bibitem{SD-d-wave-1} J. Annett, N. Goldenfeld, and S.R. Renn, Phys. Rev. B
{\bf 43}, 2778 (1991).

\bibitem{SD-d-wave-2} C. Panagopoulos and T. Xiang, Phys. Rev. Lett. {\bf
81}, 2336 (1998).

\bibitem{footnote-PPS} When the particle density $n$ is evaluated
by including fluctuation corrections (over and above the BCS contribution) to
the diagonal single-particle Green's function, (one-half of) the
(fermionic) particle density
is no longer identified with the condensate density
in the strong-coupling limit,
but rather with the sum of the (bosonic) condensate and noncondensate
densities [P. Pieri, L. Pisani, and G.C. Strinati
(unpublished)].

\bibitem{Beliaev} S. T. Beliaev, Sov. Phys.-JETP {\bf 7}, 299 (1958).

\bibitem{Popov} V.N. Popov, \emph{Functional Integrals and Collective
Excitations\/}
                (Cambridge University Press, Cambridge, 1987).


\bibitem{MPS} M. Marini, F. Pistolesi, and G.C. Strinati, Eur. Phys. J.
B {\bf 1}, 151 (1998).

\bibitem{Micnas-92} T. Kostyrko and R. Micnas, Phys. Rev. B {\bf 46}, 11025
(1992).

\bibitem{MgB2} C. Panagopoulos \emph{et al.\/}, Phys. Rev. B {\bf 64},
094514 (2001).

\bibitem{plasmon} L. Benfatto, S. Caprara, C. Castellani, A. Paramekanti,
and M. Randeria, Phys. Rev. B {\bf 63}, 174513 (2001);
L. Benfatto, A. Toschi, S. Caprara, and C. Castellani
       Phys. Rev. B {\bf 64}, 140506 (2001).

\bibitem{Levin-rho} Q. Chen, I. Kosztin, B. Jank\'{o}, and K. Levin, Phys.
Rev. Lett. {\bf 81}, 4708 (1998).

\bibitem{Bassani-01} The bosonic current correlation function within the
Bogoliubov approximation can alternatively be obtained by exploiting its 
Lehmann representation, for which the relevant matrix elements of the current 
operator can be explicitly constructed (cf. Ref.~\onlinecite{Bassani-GCS}).

\bibitem{DP-T} F. De Pasquale and E. Tabet, Ann. Phys. {\bf 51}, 223 (1969).

\bibitem{footnote-1} It has been assumed that the bosonic interaction
potential $V_B(q)$ is approximately
constant in the range $0 \leq q  \lesssim
\xi_{{\rm phase}}^{-1}$, so that $V_B(q)$ remains positive in that range.

\bibitem{Schafroth} M.R. Schafroth, Phys. Rev. B {\bf 100}, 463 (1955).

\bibitem{footnote-2} $K(\infty)$ becomes negative when
$n_{B}\xi_{{\rm phase}}^{3} < (36 \pi^{2})^{-1}
                     \simeq 2.8 \times 10^{-3}$, thus violating the
criterion for the occurrence of the Meissner effect.
                     From the relation $n_{B}\xi_{{\rm phase}}^{3} = (16 \pi
n_{B}^{1/3} a_{B})^{-3/2}$ (obtained from Eq.~(\ref{xi-phase}) with 
$n_{0}=n_{B}$ and $V_B(0)=4 \pi a_{B}/m_{B}$,
                     $a_{B}$ being the bosonic scattering length), the
above upper value for $n_{B}\xi_{{\rm phase}}^{3}$
                     is seen to coincide with the value at which the gas
parameter $n_{B}^{1/3} a_{B}$ becomes unity,
                     thus violating the diluteness condition.


\end{references}
\end{document}